\documentclass[twocolumn]{aa}
\usepackage{natbib}
\usepackage{color}
\usepackage{ragged2e}
\usepackage[pdfpagelabels=false]{hyperref}	% Hyperlinks
\hypersetup{colorlinks=true,linkcolor=blue,citecolor=blue,filecolor=blue,urlcolor=blue,}
\usepackage[varg]{txfonts}
\usepackage{graphicx,rotating}
\usepackage[normalem]{ulem}
\usepackage{colortbl}

\definecolor{darkblue}{rgb}{0.0,0.0,0.8}
\definecolor{darkred}{rgb}{0.8,0.0,0.0}
\definecolor{darkorange}{rgb}{1,0.4,0.0}
\definecolor{darkgreen}{rgb}{0.0,0.6,0.0}
\definecolor{darkpurple}{rgb}{0.8,0.5,0.9}
\definecolor{brown}{rgb}{0.65,.16,0.16}
\definecolor{grey}{rgb}{0.4,0.5,0.6}
\definecolor{trolleygrey}{rgb}{0.5, 0.5, 0.5}
\definecolor{lavender}{rgb}{0.745,0.847,0.929}

\newcommand{\mpo}{{\sc MAMPOSSt-PM}}
\newcommand{\balrogo}{{\sc BALRoGO}}

\newcommand{\gaia}{{\it Gaia}}

\newcommand{\msun}{\rm M_\odot}

\bibpunct{(}{)}{;}{a}{}{,} % to follow the A&A style

\begin{document}

\title{Properties of globular clusters formed in dark matter mini-halos}
% \title{Do globular clusters contain intermediate mass black holes? \\ A case study of \ngc\ with HST and {\sc Gaia} PMs}
\titlerunning{Globular clusters and dark matter}
\author{Eduardo Vitral\inst{1}
\and Pierre Boldrini\inst{1,2}
}
\offprints{Eduardo Vitral, \email{vitral@iap.fr}}
\institute{
$^{1}$ Sorbonne Universit\'e, CNRS, UMR 7095, Institut d'Astrophysique de Paris, 98 bis bd Arago, 75014 Paris, France \\
$^{2}$ Universit\'e de Lorraine, CNRS, Inria, LORIA, F-54000 Nancy, France
}

\authorrunning{Vitral \& Boldrini}
\date{Accepted}

\abstract{
We seek to differentiate dynamical and morphological attributes between globular clusters (GCs) that were formed inside their own dark matter (DM) mini-halo from those who were not. We employ high-resolution full $N$-body simulations on a Graphics Processing Unit (GPU) of the GCs with and without a DM mini-halo, orbiting a Fornax-like dwarf galaxy. For GCs with DM, we observed that this dark extra mass triggers a tidal radius growth that allows the mini-halo to act as a protective shield against tidal stripping, being itself stripped beforehand. We demonstrate that this shielding effect becomes negligible when the tidal radius is smaller than the half-mass radius of the mini-halo. Contrary to previous predictions, we found that the inflation of outer stellar velocity dispersion profiles is expected for GCs with and without a mini-halo, as a result of the host's tidal field. Moreover, we observed that GCs with a DM mini-halo should have, in general, relatively more radial outer velocity anisotropy profiles throughout all of their orbits, smaller degrees of internal rotation, and as a consequence of the latter, smaller ellipticities for their stellar distribution. Due to dynamical friction, we observed a clear bimodal evolutionary distribution of GCs with and without DM in the integrals of motion space and show that for GCs originally embedded in DM, this method is not reliable for association with previous accretion events. Finally, we provide parametric mass profiles of disrupted DM mini-halos from GCs that are to be used in Jeans modelling and orbital integration studies.  
}

\keywords{
(Galaxy:) globular clusters: general -- (cosmology:) dark matter -- galaxies: kinematics and dynamics -- galaxies: formation -- galaxies: dwarf -- methods: numerical
}
\maketitle

%%%%%%%%%%%%%%%%%%%%%%%%%%%%%%%%%%%%%%%%%%%%%%%%%%

%%%%%%%%%%%%%%%%% BODY OF PAPER %%%%%%%%%%%%%%%%%%

\section{Introduction}

There is no doubt that the nature of dark matter (DM) is one of the most elusive concepts in modern-day physics. However, the existence of this astrophysical component has been used and requested to explain a vast range of phenomena for a considerable amount of time, including the following: Back when \cite{Zwicky33,Zwicky37} proposed that some sort of non-luminous matter could compose the amount of mass needed to explain the discrepancy between mass measurements of the Coma cluster based on the Virial theorem (e.g. \citealt{Binney2008}) and based on the brightness and number of galaxies, up to recent measurements from the \cite{PlanckCollaboration+16,PlanckCollaboration+20} which yield very robust fits of the cosmic microwave background (CMB) using a $\Lambda$-cold dark matter ($\Lambda$CDM) model accounting for the existence of DM. For the period in between, other important confirmations of this mysterious dark component were provided: \cite{Rubin&Ford&FordKent70} and \cite*{Rubin+80} showed that the rotation curves of outer stars in nearby galaxies needed an extra amount of mass (compared to observable, luminous matter) to explain their high velocity values and finally, gravitational lensing studies (e.g. \citealt{Taylor+98}) have also confirmed that the total amount of mass in many galaxy clusters corresponds to the dynamical measurements accounting for DM. 

Such findings point to this DM as a fundamental component in galaxy formation, which is present in most galaxies as an enveloping halo, from the smallest to the highest scales. Thus, in principle, it seems curious that dense collections of stars such as globular star clusters (GCs), that span from $\sim10^5 - 10^7$~M$_{\odot}$, with some of them thought to be accreted dwarf galaxies (e.g. \citealt{Majewski+00,Bekki&Freeman03,Pechetti+22}), do not seem to require any significant amount of DM to explain their dynamical mass 
% \ev{Use better references} 
(e.g. \citealt*{Shin+13,Conroy2011};\citealt{Ibata+13,Moore1996,Baumgardt2009,Lane2010,Feng2012,Hurst2015}), although some recent works seem to suggest otherwise for a few particular sources \citep{Carlberg&Grillmair22,Errani+22}. In fact, \cite{Peebles84} proposed a formation scenario where GCs are formed inside their own DM mini-halo\footnote{Those DM mini-halos could have between $10^6 - 10^8$~M$_{\odot}$, such as general DM sub-structures or sub-halos \citep{Zavala&Frenk19}.}, and further studies defended that if formed before the end of re-ionisation\footnote{Such early formation is consistent with GCs having typical ages up to $\sim 13$~Gyr \citep{MarinFranch+09}, as well as with recent findings of high-redshift GCs by the \textit{James Webb Space Telescope} \citep{Mowla+22}.}, GCs could be smaller counterparts of galaxies (e.g. \citealt{Bromm2002}, Figure~2 from \citealt{Mamon+12}, and \citealt{Silk&Mamon12} for a review on galaxy formation).
% \pbb{C'est quoi le rapport avec les papiers de Silk et Mamon?} 

On the other hand, different formation scenarios, where GCs are not necessarily embedded in DM mini-halos also exist. For instance, GCs could be formed as bound gas clouds \citep{Peebles&Dicke68}, as galaxy DM-free fragments (e.g. \citealt*{Searle&Zinn78,Abadi+06}), as relics of young massive clusters (YMCs, \citealt*{PortegiesZwart+10,Longmore+14}) that formed in the high-redshift Universe \citep{KruijssenDiederik14,KruijssenDiederik15}, or simply that formed in situ along with its host galaxy (e.g. \citealt{Harris91,Forbes+97}). Moreover, recent cosmological simulations indicate realistic mechanisms through which all these scenarios (i.e. DM and DM-free) can actually happen (\citealt*{Trenti+15}; \citealt{Kimm+16}; \citealt*{Ricotti+16}; \citealt{Keller+20,Ma+20}), making it reasonable to argue that GCs likely originate from more than a single formation channel.\footnote{Observational evidence for different GC populations has also been provided, for instance, by noticing significant colour gradients on the GC population from different galaxies (e.g. \citealt{Cohen+98,Harris09,Wu+22}).} 

Another reason that makes the understanding and further confirmation of different formation channels difficult is that much of the main consequences of these channels are better observable in the outskirts of GCs, where one often lacks good quality data. For example, the detection of tidal tails or stellar streams that might relate to an accretion event can be complicated by an observational bias where the stars in the stream are less luminous than the ones in the central and dense regions of the cluster, and thus more difficult to observe, comparatively \citep{Balbinot&Gieles&Gieles18}. Similarly, a potential DM mini-halo could present a much more diffuse structure than the stellar component, so that its dynamical detectability might only be possible beyond several scale radii \citep{Penarrubia+17}, where GC stars are usually confused with galactic field stars.

With the astrometric revolution brought by the \gaia\ mission \citep{GaiaCollaboration+18H,GaiaCollaboration+18B,GaiaCollaboration+21}, and the promising future discoveries of the \textit{James Webb Space Telescope} (\textit{JWST}, \citealt{Gardner+06}) and the \textit{Euclid} mission \citep{Laureijs+11,Lanccon+21}, the need to better constrain the expected differences between the multiple GC formation scenarios is a priority, so that these rich data sets can be fully exploited to better understand the many long-sought questions regarding GC formation. As a matter of fact, although many robust attempts to better model the observational implications of the DM mini-halo scenario have been made, the high computational cost of simulating a GC$+$DM system in a Milky Way (MW) type of galaxy forced these attempts to be placed in idealised scenarios: for instance, isolated GCs not experiencing tidal forces \citep{Penarrubia+17}, or orbiting GCs in a static potential \citep{Mashchenko&Sills&Sills05}. Preferably, simulations with clusters experiencing tidal forces, in a host galaxy fully composed of particles (and not just a static potential) would allow to take into account more correctly dynamical friction, tidal effects between globular clusters and their host galaxy, and dynamical shocks with larger structures of the host.

In this work, we aim to clearly separate the observational behaviours of GCs that are formed inside a DM mini-halo and which are devoid of one, both of them orbiting a host galaxy and thus permanently experiencing a tidal field. We do it by performing $N$-body simulations of a GC system with and without a DM embedding mini-halo, alongside with a host galaxy. 
In order to bypass the high computational cost mentioned above while still keeping a high resolution, we place our GCs in a dwarf spheroidal (dSph) galaxy, following the same prescriptions as the Fornax dSph, in a similar manner than done in \cite*{Boldrini2020}.\footnote{Indeed, \cite{Boldrini2020} have shown that if the GCs from the Fornax dSph were formed in DM mini-halos, this could account for a natural explanation of the cusp-core and timing problems in this dwarf, adding robustness to this scenario.}. Such setup allows one to consider much less stars than it would be needed if the satellites were orbiting a MW-like galaxy and keep a satisfactory resolution, and also to avoid using a potential instead of particles. 
% \pbb{to reach the same resolution?} 
Besides, by taking a real galaxy (i.e. Fornax) as our model, we are likely better exploring the dynamics and orbital evolution of different parameters, and thus reaching more realistic conclusions. 

We describe our methods in Section~\ref{sec: method}, while our results and main DM signatures are presented in Section~\ref{sec: results}. We discuss the implications of our work and conclude it in Sections~\ref{sec: discussion} and \ref{sec: conclusion}.
Throughout the rest of the paper, we reference the stellar component of the system of GC stars without DM as $\star$, and the stellar component of the system of GCs formed inside DM mini-halos as $\star\bullet$. The dark matter is labelled as $\bullet$. 

\begin{table}
\centering
\renewcommand{\arraystretch}{1.0}
\tabcolsep=6pt
\footnotesize
\caption{Initial conditions of our simulations.}
\begin{tabular}{l|lrrrrrrrr}
\hline
\rowcolor{lavender}
\multicolumn{1}{c}{Object} & 
\multicolumn{1}{c}{} & 
\multicolumn{1}{c}{$M$} &
\multicolumn{1}{c}{$N$} &
\multicolumn{1}{c}{} &
\multicolumn{1}{c}{} \\
\rowcolor{lavender}
\multicolumn{1}{c}{} &
\multicolumn{1}{c}{} &
\multicolumn{1}{c}{[$10^{6}~{\rm M}_{\sun}$]} & 
\multicolumn{1}{c}{[$10^{4}$]} &
\multicolumn{1}{c}{} &
\multicolumn{1}{c}{} \\
\hline
\hline
Fornax dwarf &  \\
& DM & $1000$ & 2000 \\
& Stars & $38.2$ & 76.4 \\
\hline
% \hline
Globular clusters &  \\
& DM & $20$ & 40\\
& Stars & $1$ & 2\\
\hline
\hline
\rowcolor{lavender}
\multicolumn{1}{c}{} &
\multicolumn{1}{c}{} &
\multicolumn{1}{c}{$r$} & 
\multicolumn{1}{c}{$v_x$} & 
\multicolumn{1}{c}{$v_y$} &
\multicolumn{1}{c}{$v_z$}  \\
\rowcolor{lavender}
\multicolumn{1}{c}{} &
\multicolumn{1}{c}{} &
\multicolumn{1}{c}{[kpc]} & 
\multicolumn{1}{c}{[km/s]} & 
\multicolumn{1}{c}{[km/s]} & 
\multicolumn{1}{c}{[km/s]} \\
\hline
% &&\\
& {GC~1} & 5.32 & 13.90 & 1.31 & 14.38 \\
& {GC~2} & 2.07 & 9.43 & 19.05 & 21.42 \\
& {GC~3} & 1.95 & 37.49 & 4.55 & 6.21 \\
& {GC~4} & 2.19 & 3.51 & 3.87 & 34.00 \\
& {GC~5} & 2.05 & 19.62 & 29.80 & 15.10 \\
\hline
\end{tabular}
\parbox{\hsize}{Notes: From left to right, the upper columns give, for each component (dark matter and stars), the mass (in $10^6$~M$_\odot$) and number of particles (times $10^4$). We also provide initial positions and velocities of our globular clusters (with respect to the Fornax dwarf) in the lower columns.}
\label{tab: initial-cond}
\end{table}

\section{Methods} \label{sec: method}

\subsection{Initial conditions}

The initial conditions for the Fornax system were taken from \cite{Boldrini2020}, and although a complete description can be seen in their Section~2, we remind readers here of the most salient points, which are summarised in Table~\ref{tab: initial-cond}. We consider the scenario from that paper where the GCs were accreted recently, at $z=0.36$ (i.e. 4~Gyr ago) by Fornax. It ensures that at $z=0$, the GCs embedded in DM are still orbiting and no star clusters form in the centre of Fornax (in accordance with observations), and also that at 3 Gyr, the cluster's positions relative to Fornax are consistent with their observed projected distances.

The Fornax dSph is modelled with a stellar component following the Plummer \citep{Plummer1911} profile, with total mass of $3.82 \times 10^7$~M$_{\odot}$ \citep{deBoer&Fraser16} and Plummer scale radius of $1.038$~kpc \citep{Strigari+10}. Its DM halo follows a NFW \citep*{Navarro+96} profile with total mass $10^9$~M$_{\odot}$. Given the halo mass and
redshift, the scale radius of the Fornax DM halo was estimated from cosmological $N$-body simulations \citep{Prada+12}.

\begin{figure*}
\centering
\includegraphics[width=0.75\hsize]{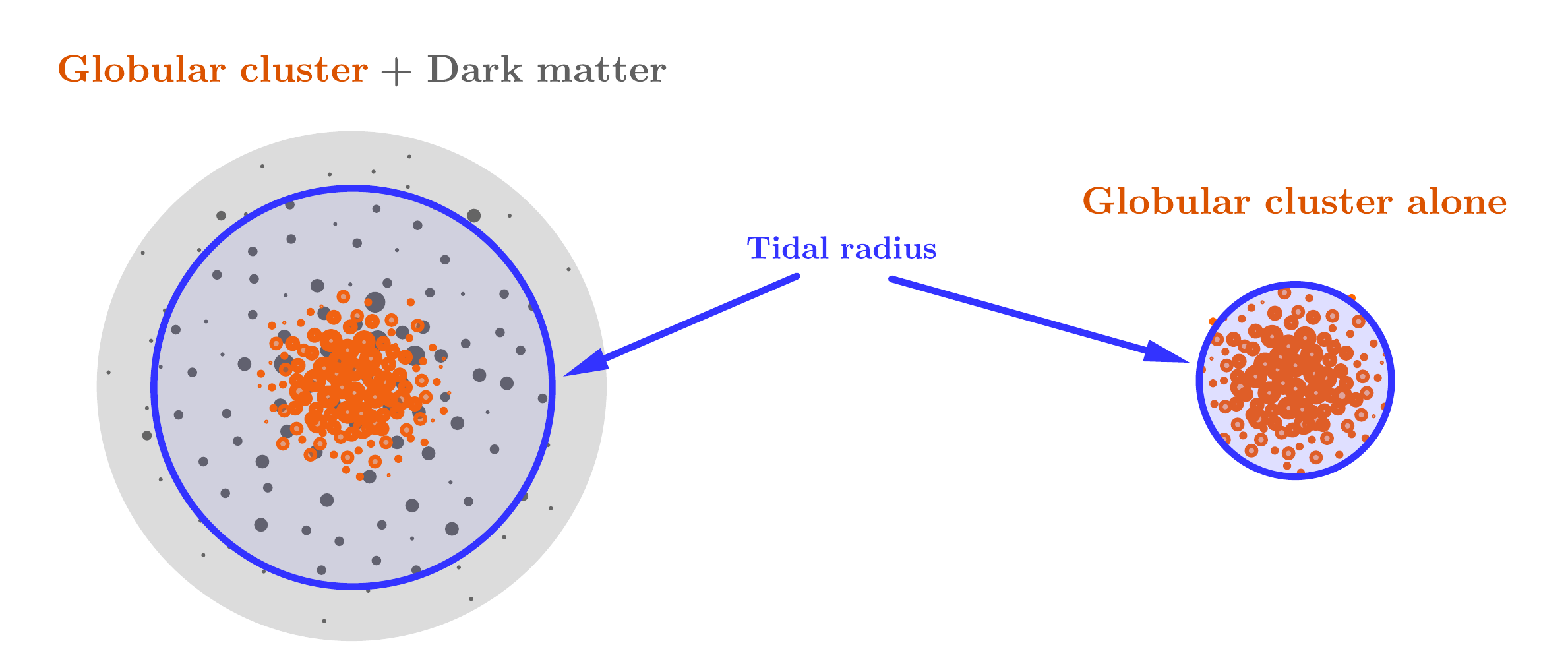}
\caption{Tidal radius growth due to an extra dark mass: Scheme illustrating the increase of tidal radius (\textit{blue}) in the globular cluster system (\textit{orange}) embedded in a dark matter mini-halo (\textit{grey}), in the \textbf{left}, in comparison with the globular cluster system without dark matter, in the \textbf{right}, at a same orbital radius. This tidal radial expansion is mostly explained by its dependence on the satellite's mass, which is greater for the case with a DM mini-halo.}
\label{fig: tidal-scheme}
\end{figure*}

The GCs on the other hand, started with a \cite{King62} stellar distribution, with total mass of $10^6$~M$_{\odot}$ and a King radius of 1~pc, lower than the observed radius, since it is susceptible to increase through dynamical processes such as mass loss \citep{Mackey&Gilmore03}. When assigning their DM mini-halo, we prescribed a mass of $2 \times 10^7$~M$_{\odot}$, also following a NFW profile, and computed the respective scale radius in a similar fashion than for the dSph.
The GC initial positions in phase space were set after analysing 7395 isolated dwarf galaxies with mass $\sim 10^9$~M$_{\odot}$ in the Illustris TNG-100 simulation \citep{Pillepich+18}, whose mass resolution is $7.5 \times 10^6$~M$_{\odot}$, hence similar to the DM mini-halo mass. The positions were then selected by picking the cases where the sub-halos projected distances to the dwarf, at $z=0$, were in the same range as the ones observed for the Fornax dSph GCs. From this subset, a new set of positions was obtained by going back 4~Gyr in time, and finally, the final values were taken by computing the maximum weights in each dimension.

Our choice of ratio for the mass of the GC mini-halo and stellar component (i.e. $M_{\rm DM} / M_{\rm stars} = 20$) is consistent with recent works targeting this scenario. For instance, \cite{Mashchenko&Sills&Sills05} simulated GCs with a stellar mass of $8.8 \times 10^4$~M$_{\odot}$ and a mini-halo virial mass of $10^7$~M$_{\odot}$, while \cite{Penarrubia+17} tested models with mass ratios $M_{\rm DM} / M_{\rm stars} = \{0,10,100\}$. Indeed, GC systems with a DM mini-halo are very similar to orbiting (or accreted) DM sub-halos, which sit in a specific mass range for dwarf galaxies, predicted by the extended Press-Schechter (EPS) formalism \citep{Bond+91,Lacey&Cole&Cole93}. 
A Fornax-like galaxy is considered to have accreted roughly twelve DM sub-halos \citep[see figure 1 from][]{Boldrini+20} with a mass ratio $M_{\rm Fornax}/M_{\rm sub-halo} \sim 10 - 10^2$ \citep{Zavala&Frenk19}, consistent with our values.
% \pbb{Tu peux citer mon papier avec les subhalos: https://arxiv.org/pdf/2003.02611.pdf. La figure 1 montre qu'une galaxie comme Fornax a accrêté environ 12 subhalos dans le mass range que tu décris}

\subsection{Simulations} \label{ssec: simus}

To generate our $N$-body objects, we use the initial condition code \textsc{magi} \citep{Miki&Umemura&Umemura18}. Adopting a distribution-function-based method, it ensures that the final realisation of the galaxy is in dynamical equilibrium \citep{Miki&Umemura&Umemura18}. We perform our simulations with the high performance collisionless $N$-body code \textsc{gothic} \citep{Miki&Umemura&Umemura17}. This gravitational octree code runs entirely on a Graphics Processing Unit (GPU) and is accelerated by the use of hierarchical time steps in which a group of particles has the same time step \citep{Miki&Umemura&Umemura17}. In order for the simulated GCs to fully relax prior to introducing the Fornax tidal field, they were previously left to evolve in isolation for 2 Gyr.

We evolve the Fornax-GC system over 4 Gyr in each scenario, i.e. for GCs with and without a DM mini-halo. We set the particle resolution of all the live objects to 50~M$_{\sun}$\footnote{We note that \cite{Banik&Bovy&Bovy21} state that mass resolutions of 100~M$_{\odot}$ or less would be needed to avoid spurious numerical density variations in stellar streams, thus well in agreement with our mass resolution.} and the gravitational softening length to 0.1~pc. Numerical convergence tests have been performed in \cite{Boldrini2020}.
The use of a host composed of live particles rather than just a static potential provides a better handling of dynamical friction, which is the deceleration of massive particles (here, the GC) after interacting with less massive particles (here, the stellar and DM particles from the host). This is necessary to interpret the evolution of orbital radii of GCs, both with and without a DM mini-halo, and to robustly quantify their differences. Furthermore, the pruning of tidal structures \citep[][Section~7.3]{Binney2008} due to the shocking of such structures with the host field tracers is an effect completely neglected when using static potentials. Hence, the study of stellar stream's properties and projected ellipticity are also better addressed when using live particles, such as we do. 
% \pbb{Bien expliqué!}

\subsection{Data analysis} \label{ssec: data-analysis}

Using the instantaneous orbital radius $r_{\rm orbit}$ and the instantaneous satellite mass $M_{\mathrm{\rm sat}}$, we calculate the \textit{theoretical} tidal radius of each GC (or GC$+$DM system) at each snapshot from the simulation as derived by \cite{Bertin&Varri08}, and used in many GC-related studies \citep[e.g.][]{Khalaj&Baumgardt16,Daniel+17,Webb+19}.
For the system composed of GC stars plus DM particles (i.e. GC$+$DM), we measure the total tidal radius in our simulations by taking into account the mass of both DM and stars in our calculation. 
In order to determine the bound particles of the DM and stellar components of GCs, we follow the procedure described in \citeauthor{vandenBosch&Ogiya&Ogiya18} (\citeyear{vandenBosch&Ogiya&Ogiya18}, see their section~2.3). The bound mass is determined by computing iteratively the binding energy of each particle. Initially, the method assumes that each particle is bound. We use the publicly available code \texttt{galpy}\footnote{Available at \url{https://github.com/jobovy/galpy}} \citep{2015ApJS..216...29B} to compute the quantities $E$ and $L_{\rm Z}$, for the integrals of motion space.

\begin{figure*}
\centering
\includegraphics[width=0.99\hsize]{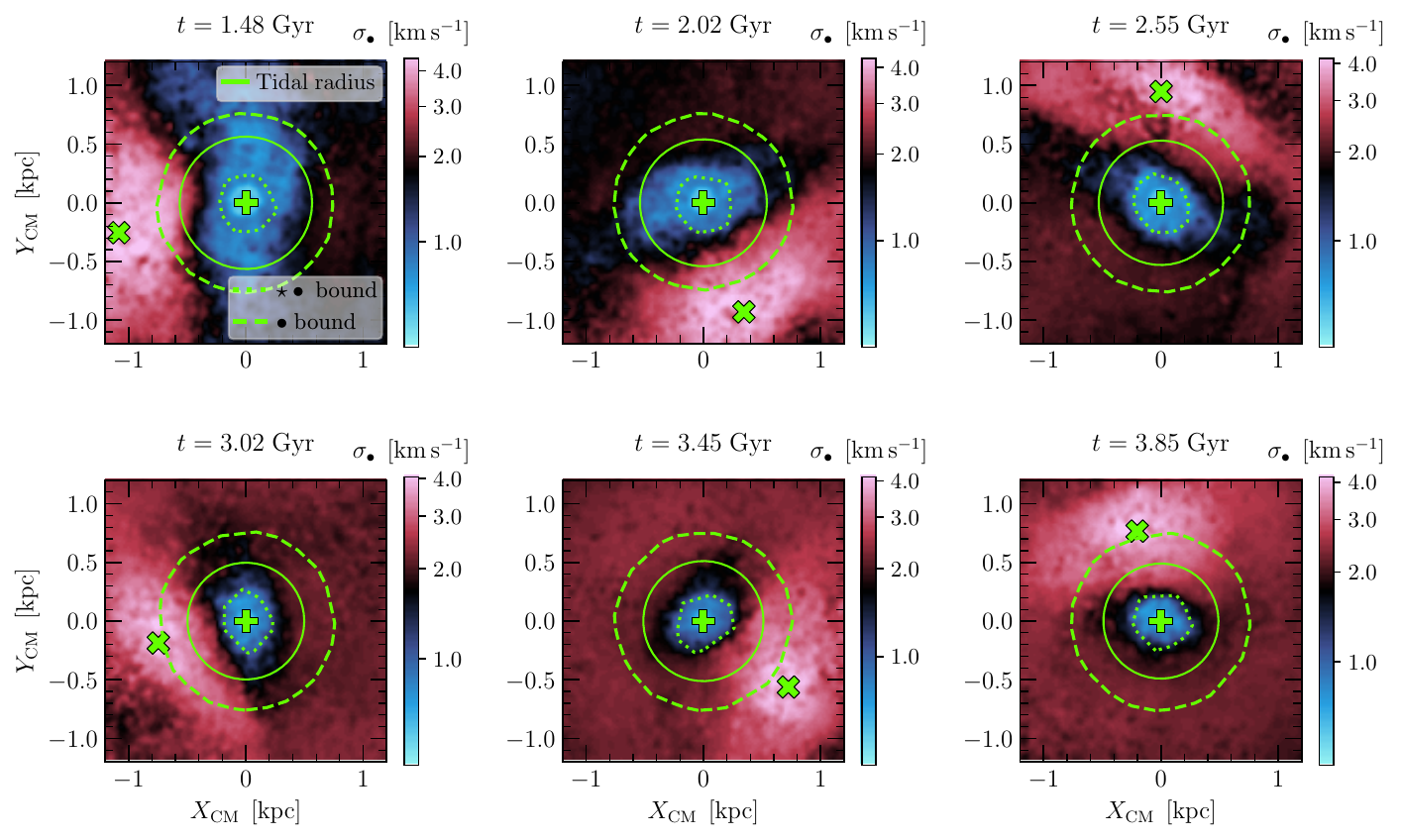}
\caption{Dark matter shield: Velocity dispersion map of dark matter particles for GC5, projected in the $X$~vs.~$Y$ plane and centred in the centre of mass of the globular cluster system. We display the last six pericentres of its orbit, where the tidal effects are stronger. The extension of bound globular cluster stars and bound dark matter particles are highlighted as dotted and dashed green lines, respectively, while the theoretical tidal radius, calculated according to section~\ref{ssec: data-analysis}, is displayed as a solid green circle. The maps are colour-coded logarithmically from blue (lower dispersion) to red (higher dispersion). The centres of Fornax and of the globular cluster are represented as a thick green cross and a plus sign, respectively. For this cluster, we notice that the empirical tidal radius, well traced by the blue region, remains always larger than the bound stars radii. This argues in favour of the dynamical presence of a dark matter shield, as illustrated in Figure~\ref{fig: tidal-scheme}.}
\label{fig: dm-shield-gc5}
\end{figure*}

We performed most of the data analysis with internal routines of the astrometry modelling code \balrogo\footnote{Code repository: \\ \url{https://gitlab.com/eduardo-vitral/balrogo}.} \citep{Vitral21}. This includes fits of the surface density (and ellipticity), construction of velocity anisotropy and dispersion profiles, and sky projection. Whenever targeting these themes, we mention in detail how we proceeded to perform the analysis.

\section{Results: Dark matter signatures} \label{sec: results}

In this section, we analyse the main implications of the presence of a DM mini-halo in the overall GC dynamics and morphology, according to our simulations. We remind that our analysis is modelled on the specific case of the Fornax dSph, in order to probe a realistic scenario.

\subsection{Dark matter shield} \label{ssec: dm-shield}

In the beginning of our simulations with DM mini-halos, all GCs presented a DM envelope massive and concentrated enough, so their total tidal radius was considerably greater than the case without a DM mini-halo (see Figure~\ref{fig: tidal-scheme}). 
Whenever we had such tidal radius increase, the DM particles are stripped beforehand GC stars, so the DM envelope works effectively as a shield against tidal stripping of the stellar component, being gradually removed as the system experiences stronger tidal forces from the host galaxy.

In order to visualise this effect, we created a velocity dispersion map (see appendix~\ref{app: disp-map} for details) of DM particles, which are much more spread than GC stars, and thus provide better spatial completeness to the map. Such map helps us to spot the transitory region where the DM particles start to effectively feel tidal effects of the host galaxy, which is characterised by a steep increase of the velocity dispersion produced by tidal heating of the system.

In Figure~\ref{fig: dm-shield-gc5}, we show this velocity dispersion map for GC5, one of the clusters where this DM shield seemed most effective. We display the last six pericentres of its orbit, when the tidal effects are stronger. The extension of bound GC stars and bound DM particles are highlighted as dotted and dashed green lines, respectively, while the theoretical tidal radius, calculated according to section~\ref{ssec: data-analysis}, is displayed as a solid green circle. The centres of Fornax and of the GC are represented as a thick green cross and a plus sign, respectively.

We also observe, still in Figure~\ref{fig: dm-shield-gc5}, a gradual decrease in the extension of the blue region, characterised by a low velocity dispersion, where the DM shield is effective. This means that at first, the DM shield is highly protective, and with time, as DM particles are stripped from the cluster, the system mass decreases, and the shield becomes weaker.
The red colours, on the other hand, point to a region of high dynamical heating, which becomes more intense in the centre of Fornax and favours tidal stripping of the DM shield.

\begin{figure*}
\centering
\includegraphics[width=0.99\hsize]{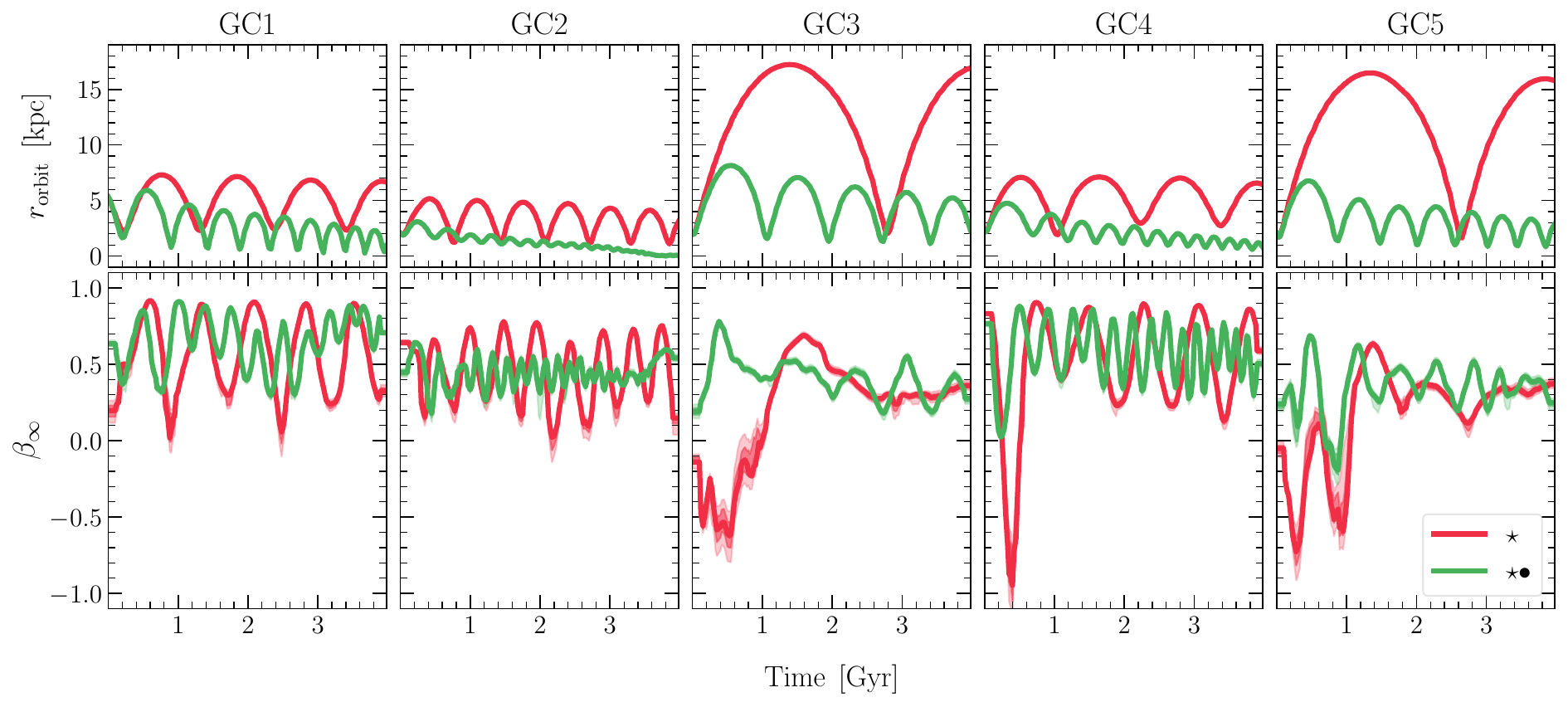}
\caption{Velocity anisotropy: Evolution of the orbital radius (i.e. $r_{\rm orbit}$, \textit{upper plots}) and of the outer velocity anisotropy (i.e. $\beta_{\infty}$, \textit{lower plots}), for the five simulated globular clusters. The clusters originally embedded in a dark matter mini-halo are depicted in \textbf{solid green}, and those without it are displayed in \textbf{solid red}. For $\beta_{\infty}$, we display the running mean over the five closest snapshots (time-wise), for better visualisation, which explains why the initial $\beta_{\infty}$ values seem to be different for the two scenarios. We observe typically lower $r_{\rm orbit}$ and more radial (i.e. higher) $\beta_{\infty}$ for clusters with a dark matter mini-halo.}
\label{fig: evolution-anis}
\end{figure*}

This protective DM shield is a key mechanism to explain most of the DM signatures we observed in GCs embedded in DM. Although this phenomenon is ubiquitous in our GCs formed in DM mini-halos, the survival of such shield depends mostly on the orbital parameters of each cluster, and a specific discussion on this matter is addressed in Section~\ref{ssec: survival-dm}. In the following, we focus on describing the main impacts of such shield with respect to the case where GCs are not embedded in DM mini-halos.

\subsection{Orbital decay} \label{ssec: decay}

The exchange of energy between the satellite and the field stars from its host galaxy will lead to the drag of the satellite, also referred to as dynamical friction. The use of live particles for both the satellite and its host, such as we do, allows to provide a better handling of this effect.
The dynamical friction time of the cluster is the timescale needed for the satellite to reach the centre of mass of the host galaxy. It has been defined in \citeauthor{Binney2008} (\citeyear{Binney2008}, eq [7-26]), and follows the relation $t_{\rm fric} \propto 1/M$, where $M$ the satellite’s total mass.

Applying this relation to our simulations, where in one case we have a GC system alone of $10^{6}$~M$_{\odot}$ and in the other case where the total system mass is that of the GC plus the DM mini-halo (i.e. $2 \times 10^{7}$~M$_{\odot}$), we find that the system with DM is supposed to sink to the centre twenty times faster than the system without the DM mini-halo. Indeed, when looking at the upper panels of Figure~\ref{fig: evolution-anis}, one can notice that the systems with DM (solid green) occupy much shorter orbital radii than the systems without DM (dashed red) throughout their orbits.

As a consequence, systems with a DM mini-halo tend to be located closer to their host centre sooner, and thus feel a stronger dynamical heating imposed by the host galaxy. This dynamical heating, on its turn, can potentially work to remove more GC stars (along with their DM envelope) than in the case without DM. Hence, our simulations help to answer whether the presence of a DM mini-halo is rather protective or disruptive with respect to the cluster's stellar component, a point that we discuss in detail in Section~\ref{ssec: survival-dm}.

\subsection{Velocity anisotropy} \label{ssec: anis}

One of the points not addressed in previous works targeting a DM mini-halo in GCs \citep[e.g.][]{Mashchenko&Sills&Sills05,Penarrubia+17} is the evolution of the velocity anisotropy of the stellar component, in both cases with and without a DM mini-halo. The velocity anisotropy (`anisotropy' for short) is defined as in \citep{Binney80}:
\begin{equation}    \label{eq: anisotropy}
    \beta(r) = 1 - \displaystyle{\frac{\sigma_{\theta}^{2}(r) + \sigma_{\phi}^{2}(r)}{2 \,\sigma_{r}^{2}(r)}} \ ,
\end{equation}
where $\theta$ and $\phi$ are the tangential components of the coordinate system, while $\sigma_{i}^2$ stands for the velocity dispersion of the component $i$ of the coordinate system. In spherical symmetry, $\sigma_\phi = \sigma_\theta$. 

We fit the anisotropy for each cluster and snapshot of our simulations in a Bayesian frame detailed in Appendix~\ref{app: anis}. For that, we consider the generalised parametrisation from \cite{Osipkov79,Merritt85}, which has been applied in many works modelling the dynamics of spherical stellar systems \citep[e.g.][]{Walker&Penarrubia&Penarrubia11,Mamon+13,Read+21,Vitral&Mamon21,Vitral+22}
\begin{equation}    \label{eq: gOM}
    \beta_{\mathrm{gOM}}(r) = \beta_{0} + (\beta_{\infty} - \beta_{0}) \ \displaystyle{\frac{r^2}{r^2 + r_{\beta}^2}} \ ,
\end{equation}
where $r_{\beta}$ is the anisotropy radius, and $(\beta_0,\beta_{\infty})$ are the anisotropy values at the centre of the system, and at infinity, respectively. We keep in mind that the initial $\beta(r)$ in our simulations is not necessarily isotropic as a result of the prior 2 Gyr relaxation run we mention in Section~\ref{ssec: simus}, which can change the anisotropy shape, specially in the cluster outskirts ($\beta_0$ still lies closer to zero).

\begin{figure*}
\centering
\includegraphics[width=0.95\hsize]{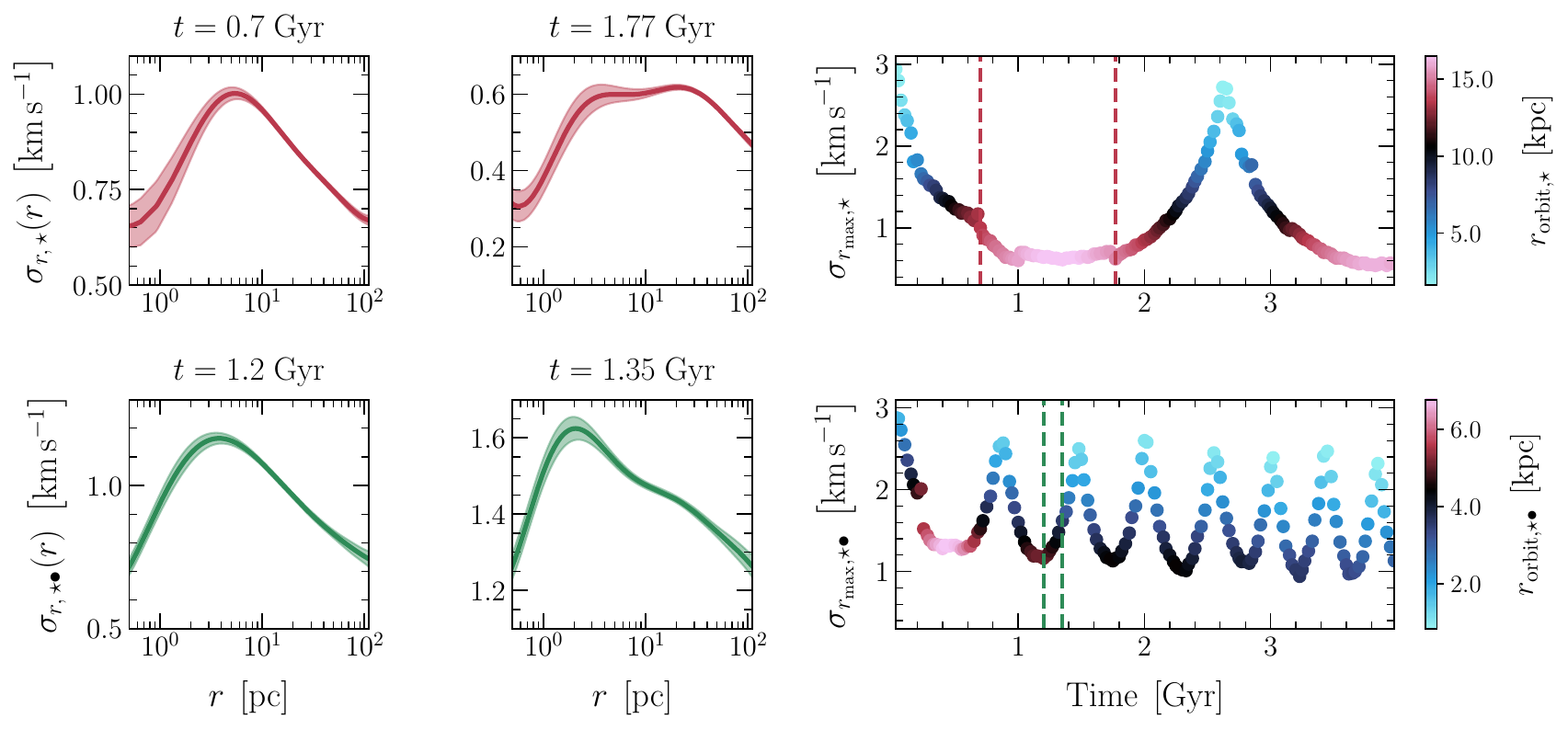}
\caption{Stellar velocity dispersion: Series of plots for GC5. The \textbf{upper} plots relate to the simulations devoid of dark matter mini-halos ($\star$), while the \textbf{lower} plots indicate the results for the globular clusters formed in such mini-halos ($\star\bullet$). The two columns on the \textbf{left} display hand-picked snapshots where the radial velocity dispersion profile (i.e, $\sigma_{r}(r)$) resembled better to an isolated case without a dark matter mini-halo (left), and with a massive dark matter mini-halo (right), according to Figure~2 from \protect\cite{Penarrubia+17}. The column on the \textbf{right} presents the evolution of the maximum value of the radial velocity dispersion profile (i.e, $\sigma_{r_{\rm max}}$), for each scenario concerning the dark matter mini-halo, colour-coded according to the distance of the cluster to the centre of the host galaxy (i.e. $r_{\rm orbit}$), with two vertical dashed lines corresponding to the instants from the two columns on the left. These plots argue that the tidal field from the host galaxy tends to have a much greater impact on inflating the velocity dispersion than the presence of a dark matter mini-halo. In fact, such mini-halos help to \textit{protect} the cluster from tidal effects, rather than contributing to it.}
\label{fig: evolution-dmax}
\end{figure*}

As we are most interested in the differences pertaining to the external dynamics of the GC systems, we chose to plot in Figure~\ref{fig: evolution-anis} (\textit{lower plots}) the evolution\footnote{We display the running
mean over the five closest snapshots (time-wise), for better visualisation, which explains why the initial $\beta_{\infty}$ values seem to be different for the two scenarios.} of $\beta_{\infty}$ for the GCs with a DM mini-halo (solid green) and without it (solid red). This allows us to quantitatively differentiate the impact of the Fornax tidal field on the cluster outer regions. 
Indeed, the external tidal field has been shown to play a major role on the outer anisotropy of GCs, by removing stars in radial orbits in the cluster's outskirts, which are more easily pruned by tides \citep[e.g.][]{Takahashi95,Tiongco+16,Zocchi+16,Bianchini+17}. As a result, one should expect more tangential (or equivalently, less radial) anisotropy profiles in systems that undergo more severe tidal interactions.

The variations of $\beta_{\infty}$ are more intense for clusters without the DM mini-halo, since they cover a broader range of orbital radii. Furthermore, an important diagnosis is that the outer anisotropy of clusters with the DM mini-halo is, in general, more radial, even though those clusters occupy regions closer to the host galaxy. We thus associate this tidal resilience with the dynamical presence of the DM shield discussed in Section~\ref{ssec: dm-shield}.

We also notice that the initial value of $\beta_{\infty}$ (not necessarily zero) does not really impact these conclusions. This is because the values of $\beta_{\infty}$ not only depend considerably on the orbital radius, but their variation speed is almost identical to the respective orbital radius variation. This means that regardless of the initial anisotropy conditions, the values of $\beta_{\infty}$ adapt rapidly to the ongoing tidal field and its further interpretation is more straightforward.

Hence, the outer anisotropy is an important parameter that allows to differentiate both scenarios. While clusters with more radial orbits (i.e. $\beta_{\infty} \gtrsim 0.4$) could belong to both scenarios, clusters with tangential anisotropy, or less radial orbits (i.e. $\beta_{\infty} \lesssim 0.4$)  were mostly obtained in our simulations without a DM mini-halo. This does not necessarily mean that clusters with very tangential orbits did not have a DM mini-halo in the past, but rather that searches for a current mini-halo surrounding them would likely be in vain. 
% \pbb{Très bien ca}

\subsection{Velocity dispersion profile} \label{ssec: dispersion-stars}

\cite{Penarrubia+17} simulated isolated GCs embedded in DM mini-halos and showed that due to the extra DM mass, an inflation of the radial velocity dispersion profile towards outer radii\footnote{Their models presented an inflated structure at roughly $r \approx 20\, r_{1/2}$, where $r_{1/2}$ is the half-mass radius.} is to be expected. In the current work, we are able to test such predictions for a more realistic scenario, where the GC$+$DM subsystem is orbiting a host galaxy, and therefore experiencing tidal effects. For that, we measured the radial velocity dispersion profiles (see Section~\ref{app: disp-prof} for details) of the bound stars in our clusters in both the cases with and without a DM mini-halo, through the course of their evolution in the Fornax tidal field. 

In Figure~\ref{fig: evolution-dmax} (right plots), we show the values of the maximum radial velocity dispersion as a function of time, colour-coded by the distance to the centre of Fornax (i.e. $r_{\rm orbit}$), in kpc. One of the most important realisations of our analysis is the fact that the overall shape of the velocity dispersion is much more impacted by the host galaxy's tidal field than by an eventual DM mini-halo: The amplitude of the dispersion, traced by its value at the peak follows a periodic variation, with same period than the GC orbit, and has values almost uniquely dependent on the ongoing tidal forces.

In the plot, we can clearly observe that at pericentres (blueish), the velocity dispersion inflates as a whole: the tidal heating from the host galaxy is effectively felt more intensively, leading to a higher velocity dispersion. In contrast, at apocentres (reddish or blackish), the cluster is closer from an ideal isolated scenario, and tidal heating is less effective, leading to low velocity dispersion peaks.

To illustrate the much stronger dependence on tidal forces than on possible DM mini-halos, we selected two snapshots for GC5 for the case with and without a DM mini-halo (lower and upper plots, respectively). In these snapshots (Figure~\ref{fig: evolution-dmax}, four left plots), we can verify that the radial velocity dispersion profile of both the DM and DM-free scenarios assume forms similar to both the isolated cases with no DM mini-halo (left), and with a massive DM mini-halo (right), as presented in Figure~2 of \cite{Penarrubia+17}.

As a general trend, all the clusters had an increasing velocity dispersion close to pericentre, with multiple points of velocity dispersion inflation throughout the GC radial extension. In apocentres, as mentioned above, the clusters resembled better to an isolated case (with one or two inflation points), specially for the case with DM mini-halo, where the shapes retrieved by \cite{Penarrubia+17} could be better spotted. The reason behind the best resemblance in apocentre for the case with DM mini-halos is directly connected to the DM shield, which protects the GC stars and approaches better the ideal isolated framework.

\subsection{Rotation \& ellipticity (flattening)}

\begin{figure}
\centering
\includegraphics[width=0.9\hsize]{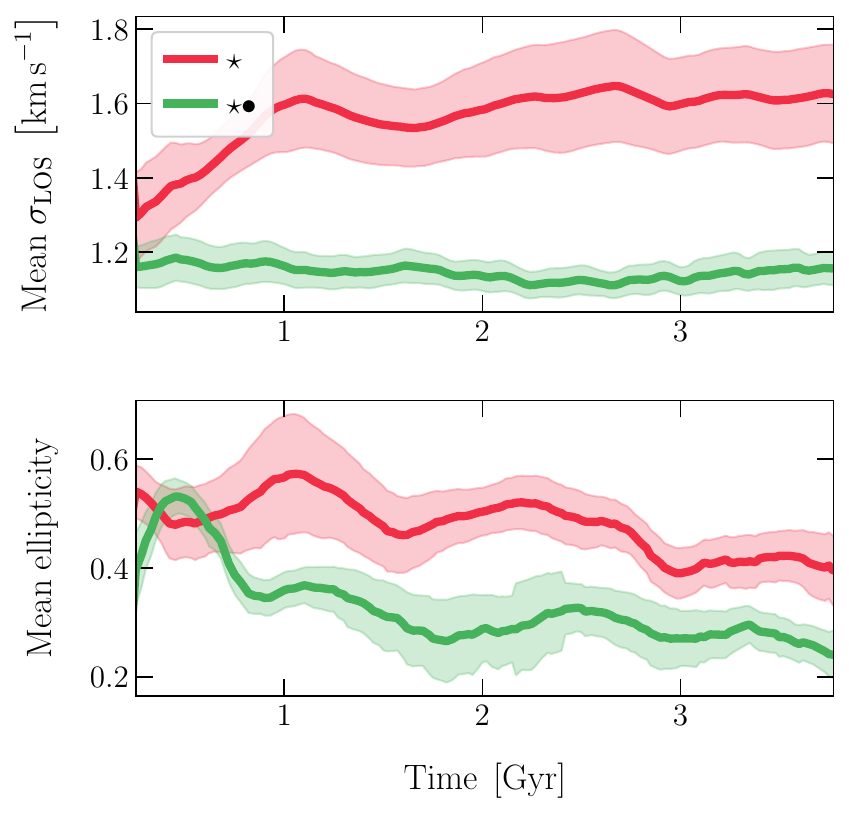}
\caption{Rotation: On the \textbf{top}, we show the mean evolution of the line-of-sight (LOS) velocity dispersion of bound globular cluster particles, for each simulated cluster, while the \textbf{bottom} plot displays the mean ellipticity (or flattening) of the stellar distribution. We display the running mean over the ten closest snapshots (time-wise), for better visualisation. Error bars are represent 1-$\sigma$ Poisson uncertainties. Clusters originally embedded in a dark matter mini-halo (i.e. $\star\bullet$) are displayed in \textit{green}, while those devoid of the mini-halo (i.e. $\star$) are represented in \textit{red}. As a general trend, globular clusters embedded in dark matter mini-halos exchange less internal angular momentum from tidal interactions, thus presenting smaller degrees of rotation, which in turn is associated with smaller ellipticities. This again attests the efficacy of the dark matter shield illustrated in Figure~\ref{fig: dm-shield-gc5}.}
\label{fig: rotation}
\end{figure}

Satellites that are more affected by tidal effects should exchange more internal angular momentum from energy transfer with their host, and this will directly impact the degrees of internal rotation of the analysed system \citep*[e.g.][]{Tiongco+16_tidal,Tiongco+18}. Hence, to quantify the differences pertaining to rotation between GCs formed in DM mini-halos and those who were not, we plot in the upper panel of Figure~\ref{fig: rotation}, the evolution of their mean line-of-sight (LOS) velocity dispersion\footnote{We display the running mean over the closest ten snapshots, for better visualisation.} (with 1-$\sigma$ Poisson error bars), since this quantity is used in many observational studies to constrain the degree of rotation of GCs \citep[e.g.][]{Bianchini+13,Boberg+17,Lanzoni+18}.
% \pbb{A c'est pas du tout évident, tu m'expliqueras le lien entre rotation et line of sight velocity please}
In order to recreate a two-dimensional projected distribution of stars, such as we observe in true data, and extract the respective LOS velocity, we placed, for each snapshot, the GC centre in the position $(\alpha,\,\delta) = (0,\,0)$, at five kpc away from the observer\footnote{Very high distances would make the stars look closer, which can decrease the precision of our fits, and also erase sky-projection signatures likely to be observed in real data. Very short distances, on the other hand, could have unrealistic sky-projection effects. We argue that 5~kpc is a good compromise, as there is a fair amount of galactic GCs with well measured data, at roughly this distance.}. The dispersion in LOS velocity of all bound GC stars is neatly higher for clusters without a DM mini-halo, meaning that they suffered more prominent tidal interactions. In other words, GCs embedded in dark matter mini-halos exchange less internal angular momentum from tidal interactions, thus presenting smaller degrees of rotation.
% \pbb{Je pense que t'as oublié de donner la conclusion pour la rotation. Il faut redire ce que tu as dit dans la caption de la figure 2.}

An increased amount of rotation can also contribute to larger values of ellipticities (or flattening) for the stellar distribution from GCs \citep{Fabricius+14,Kamann+18}. We thus fitted the distribution of bound stars with a S\'ersic \citep{Sersic63,Sersic68} asymmetric model, using the recipe described in Appendix~\ref{app: sersic}. The fits yielded a semi-major ($a$) and semi-minor axis ($b$), from which the ellipticity (or flattening) could be calculated as
\begin{equation} \label{eq: ellipse}
    e = \sqrt{1 - \left(\frac{b}{a}\right)^2} \ .
\end{equation}

The lower panel from Figure~\ref{fig: rotation} was then constructed by computing, at each instant, the mean ellipticity\footnote{Again, we display the running mean over the closest ten snapshots, for better visualisation.} of the five simulated clusters with and without DM. The error bars were calculated as $\sqrt{\left<\epsilon\right>^2 + \sigma^2_{e}/n}$, where $\left<\epsilon\right>$ is the mean uncertainty of the ellipticity fit, $\sigma_{e}$ is the standard deviation of the values from the $n=5$ clusters. 

First, we observe a decreasing pattern in the mean ellipticities of both scenarios, which is expected from orbiting satellites, due to the stripping of outer stars by the tidal field \citep{Akiyama91}\footnote{Such an effect is better modelled by the use of live particles for both the host and the satellite, as in our simulations.}. Nonetheless, one notices that clusters embedded in a DM mini-halo show considerably smaller ellipticities (i.e. $e \lesssim 0.35$) than the clusters devoid of DM, which presented rather $e \gtrsim 0.35$. Such relatively small ellipticities are indeed predictable for systems less affected by tidal forces \citep{vandenBergh08}, attesting the efficacy of the DM shield illustrated in Figure~\ref{fig: dm-shield-gc5}. 

\begin{figure*}
\centering
\includegraphics[width=0.99\hsize]{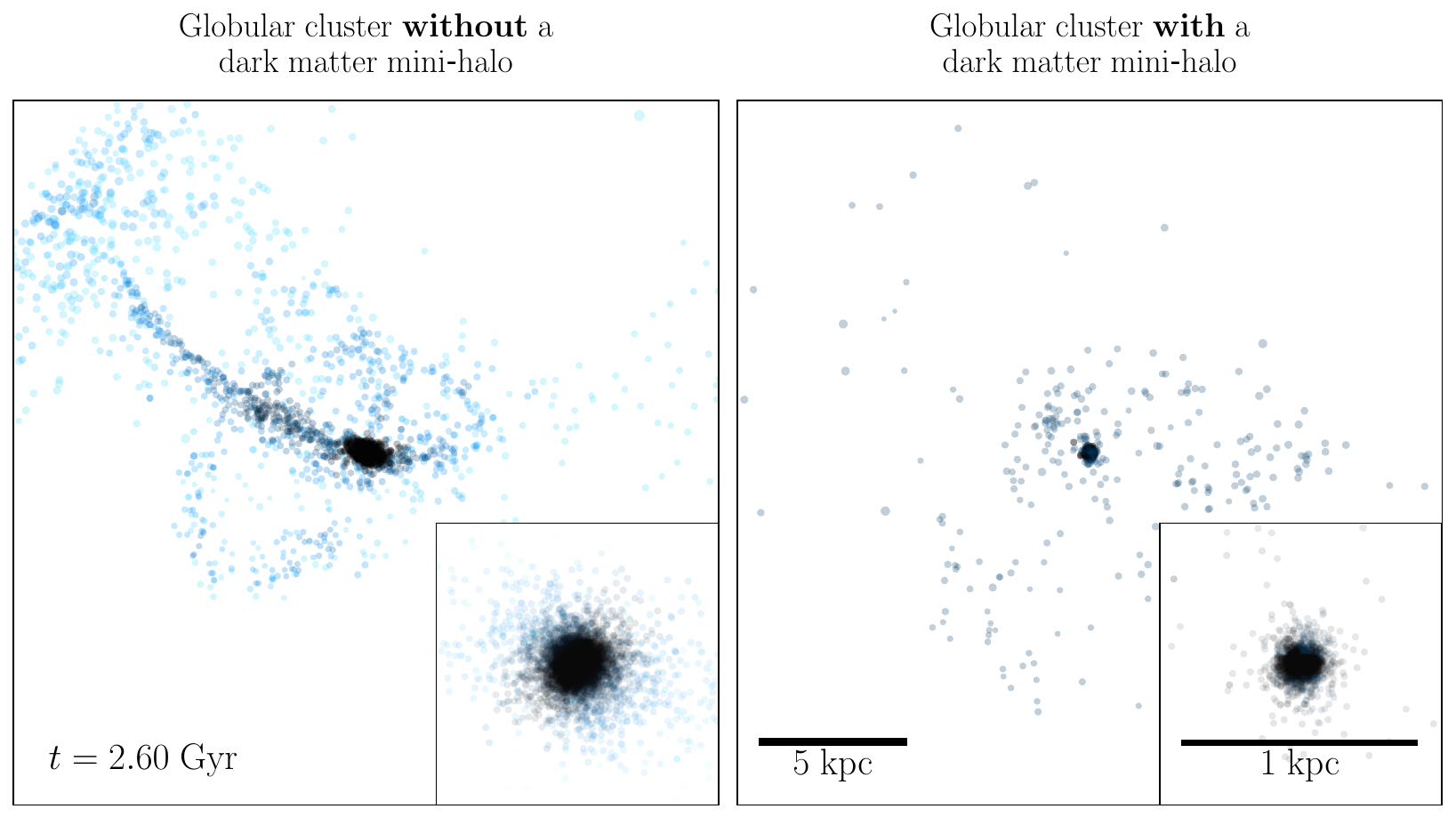}
\caption{Dark matter influence on tidal tails: Stellar distribution (and respective zoom in) of GC5 for the case where it was originally embedded in a dark matter mini-halo (\textit{right}) and where it was devoid of it (\textit{left}), both centred in the GC centre of mass. The chosen snapshot was such that GC5 was located at roughly the same orbital radius for both scenarios, for a fairer comparison. This plot highlights more prominent tails in the dark matter-free case, while clusters formed inside dark matter mini-halos present a more compact stellar envelope. Hence, the dark matter shield (Section~\ref{ssec: dm-shield}) has the effect of delaying the formation of tidal tails.}
\label{fig: tails-final}
\end{figure*}

\subsection{Tidal tails} \label{ssec: tails}

The study of tidal tails in GCs has been revolutionised by \gaia\ data and simulations. For instance, a troubling question that arises when simulating GCs on tidal fields is to understand why these simulations usually predict much more prominent tidal tails (e.g. \citealt{Boldrini&Vitral21,Montuori+07}) than what is observed for the majority of MW clusters. The answer to this timing problem is partially given by \cite{Balbinot&Gieles&Gieles18}, who showed with simulations that there is a preferential bias towards the escaping of low-mass stars, specially in denser clusters. Such trend reduces considerably the visibility of the tails. In addition, \cite{Gieles+21} recently defended this trend by showing that the visible and extended tails of Palomar~5 are well explained by a supra-massive population of stellar-mass black holes, which is a characteristic associated with less dense GCs \citep{Kremer+20}. On the other hand, the presence of a DM mini-halo could also reduce the prominence of tidal tails, since the mini-halo is expected to be stripped beforehand GC stars, thus delaying tail formation (e.g. \citealt{Bromm2002,Mashchenko&Sills&Sills05,Saitoh2006,Bekki2012,Boldrini&Vitral21}). 

In any case, the increase of better quality data such as \gaia\ EDR3 has allowed to go deeper into this question: Although some GCs, such as NGC~1851 and NGC~7089 (M2), were thought not to have tails based on ground-based imaging \citep{Kuzma+16,Kuzma+18}, further \gaia\ studies revealed long tails associated to them \citep{Ibata+21}. However, many clusters with no tidal features, or with only extended envelopes (without tails, see \citealt{Piatti&CarballoBello20}) are still present\footnote{See Table~3 from \citealt{Zhang+22}.}, and there is still no consensus on which mechanisms are behind the lack of very extended tails in these GCs. 
On a similar note, \cite{Martin+22} recently used \gaia\ EDR3 and CFHT data to constrain a stellar stream, C-19, whose metallicity is consistent with it being a remnant of the oldest GC known in our Galaxy, even though the stream still presents a coherent structure to this day, thus requiring some sort of shielding mechanism to protect it from tidal stripping.

We decide to explore the consequences of a DM mini-halo on tidal tails, and we illustrate it in Figure~\ref{fig: tails-final}, with the stellar distribution of GC5 (one of the cases where the DM shield was most effective), for the case with (\textit{right}) and without (\textit{left}) an embedding DM mini-halo. The chosen snapshot was such that GC5 was located at roughly the same orbital radius for both scenarios, for a fairer comparison. We also provide a video of the evolution of the tails in the two kind of clusters in the footnote link\footnote{Link here: \\ \url{https://gitlab.com/eduardo-vitral/vitral_boldrini/-/blob/main/movie.mp4}.}, with a respective zoomed-in version in the second footnote link\footnote{Second link here: \\ \url{https://gitlab.com/eduardo-vitral/vitral_boldrini/-/blob/main/movie_zoom.mp4}}.
We observe much more prominent and obvious tails in the case without DM, with thick and well-defined streams measuring up to $\gtrsim10$~kpc long. However, in the case where the GCs are formed inside DM mini-halos, the stellar distribution remains roughly spherical. We connect this difference between two kind of GCs to the protection of the DM shield (as explained in Section~\ref{ssec: dm-shield}, and in Figures~\ref{fig: tidal-scheme}, \ref{fig: dm-shield-gc5}), which reduces and delays tidal effects on the GCs where it is present. 

Hence, although GCs embedded in DM mini-halos can still form stellar streams eventually, the ones devoid of DM seem to develop much longer, thicker and well-defined tails. Evidently, once the DM shield is destroyed, which can happen much before a Hubble time for clusters having specific orbital parameters (see Section~\ref{ssec: survival-dm}), the development of tidal tails can be considered similar to the case without DM, and thus even clusters originally embedded in DM could present extended tails by present time, in those conditions. Having said that, the DM shield has the effect of delaying the formation of tidal tails, by making them much milder while the shield is present, and thus providing a possible answer to the timing problem mentioned in the beginning of this Section.

\begin{figure}
\centering
\includegraphics[width=0.99\hsize]{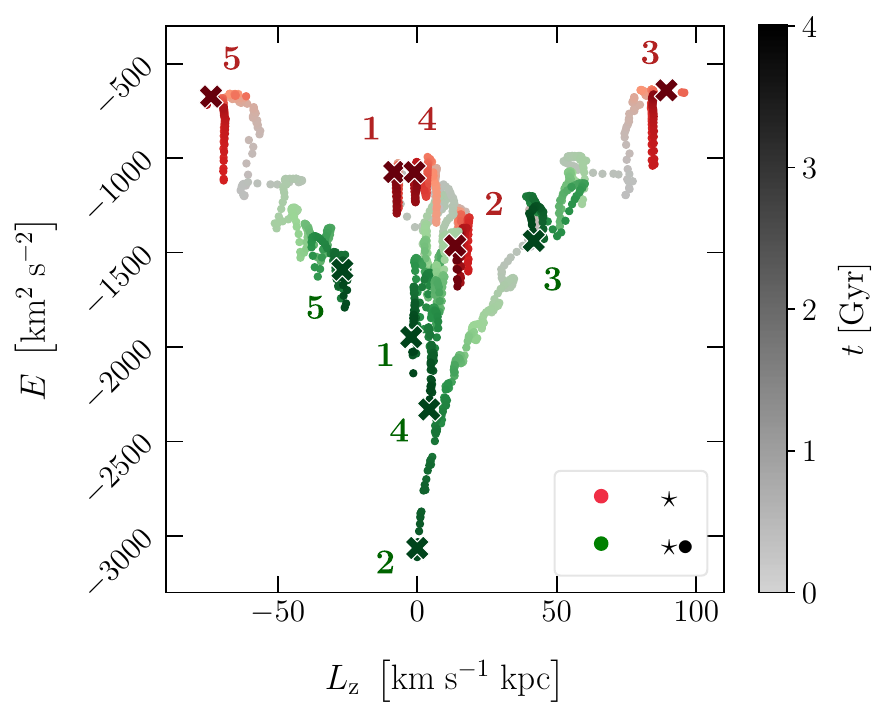}
\caption{Integrals of motion: Evolution of our simulated clusters in the integrals of motion (IOM) space, in particular, the $E \times L_{\rm Z}$ diagram. Their positions in this diagram are scattered in a sequential colour-map, starting from the same point in light grey at $t = 0$, and ending in darker tones (\textit{green} for globular clusters originally formed in dark matter mini-halos, and \textit{red} for clusters devoid of dark matter). The last snapshot is marked by a cross, for each cluster. This plot highlights a bimodal evolutionary distribution of clusters originally embedded in dark matter (moving towards lower energies) and those who were not (moving only slightly towards higher energies), and more importantly, it shows that clusters originally embedded in dark matter move significantly in IOM space, such that their association with past merger events through this diagram is not reliable.}
\label{fig: action-angle}
\end{figure}

\subsection{Integrals of motion}

By analysing the orbital evolution of simulated satellites in a static MW potential, \cite{Helmi&deZeeuw&deZeeuwTim00} proposed that the space of integrals of motion (IOM) was an optimal tool to recover the orbital history of accreted and disrupted satellites. In particular, by analysing the diagram of energy versus angular momentum in the Z galactic direction, they showed that GCs originating from an accreted satellite should remain in a similar position than its respective progenitor in IOM space, regardless of their phase and velocities being no longer the same. Even though energy is not conserved throughout orbital evolution of the satellite, the authors argued that those IOM positions should not move significantly from their original positions. Indeed, by using such method, \cite*{Massari+19} related the origin of the galactic GC population to different past major accretions events from the MW.

We decided to analyse the impact of a DM mini-halo in the evolution of a GC in IOM space by plotting their respective position in the energy-angular momentum (i.e. $E \times L_{\rm Z}$) diagram for our five analysed clusters, at each snapshot. Figure~\ref{fig: action-angle} displays this evolution for both scenarios: with a DM mini-halo in green, and without it in red, both beginning in the same position, marked by a light grey colour, and ending at their respective darker crosses. 

One observes that clusters devoid of DM (red) do not significantly move in the diagram throughout their evolution, in agreement with \cite{Helmi&deZeeuw&deZeeuwTim00}, which in turn serves as a validity check for our simulations. However, the same cannot be said for GCs originally embedded in DM, which move considerably during their orbital evolution. This is mainly due to the massive loss of energy from dynamical friction (see Section~\ref{ssec: decay}), and further changes on the host potential, which are better handled by our use of an evolving host composed of particles, instead of static potentials.
Furthermore, this energy loss shifts the GC population formed in DM mini-halos to lower regions of the $E \times L_{\rm Z}$ diagram, creating a clear bimodal evolutionary distribution in Figure~\ref{fig: action-angle}, where green evolution tracks move downwards (lower energies), while the red ones tend to move only slightly upwards (higher energies). This does not imply though, that \textit{at a given time}, a cluster with low energy necessarily had a DM mini-halo, since a DM-free GC could naturally form in such regions and not move significantly thereafter.

Given that there is no reason why these conclusions should change qualitatively when considering the case of the MW, Figure~\ref{fig: action-angle} has important implications. Not only it shines a light on the bimodal evolutionary distribution commented above, which can be used to guide the search of GCs candidates that have been formed in a DM mini-halo, but most of all, it shows that GCs formed in DM mini-halos move significantly in IOM space due to their more intense energy loss from dynamical friction. Hence, the association of such clusters with past merger events by using the IOM method is not reliable. Indeed, \cite{Boldrini&Bovy&Bovy21} recently showed, by performing orbital integrations, that there was no clear association between many MW GCs and other MW satellites, even though a large fraction of them are believed to be accreted from previous IOM analyses. 
% \pbb{Parfait!}

\begin{figure*}
\centering
\includegraphics[width=0.99\hsize]{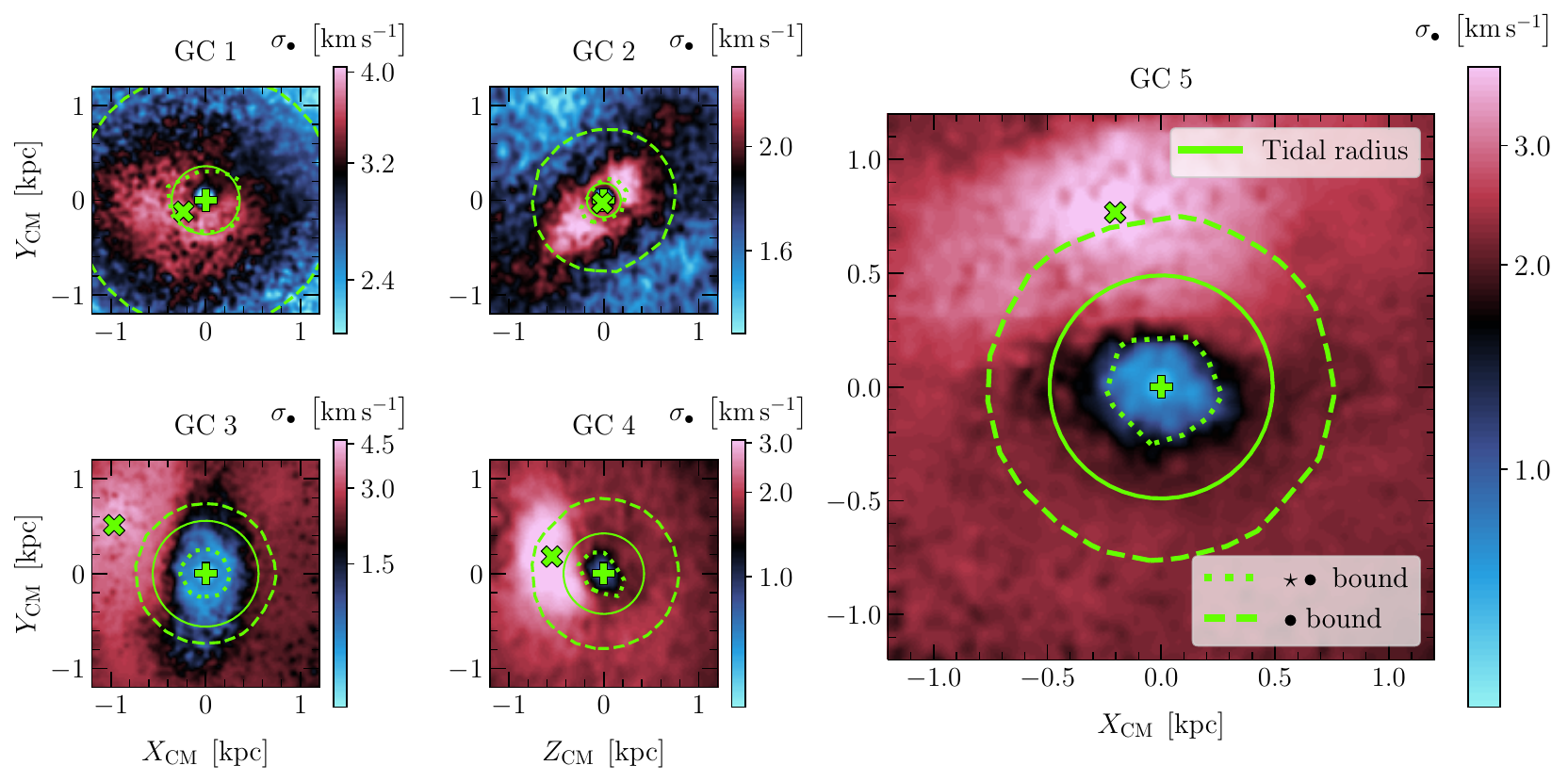}
\caption{Survival of the dark matter mini-halo: Similarly to Figure~\ref{fig: dm-shield-gc5}, we show the velocity dispersion map of dark matter particles for the five globular clusters we simulated, projected in the $X$~vs.~$Y$ plane (with exception of GC4, projected on the $Z$~vs.~$Y$ plane for better visualisation of its orbit) and centred in the centre of mass of the globular cluster system. We display the last pericentre passages of their respective orbits, where the tidal effects are stronger.
One can notice that while GCs~3, 4 and 5 manage to keep the dark matter mini-halo (and thus the dark matter shield effect depicted in Figure~\ref{fig: dm-shield-gc5}), GCs~1 and 2 do not. The extent of the red region for GCs~1 and 2 is related to their different orbital parameters.
For GCs~3, 4 and 5, we again notice that the empirical tidal radius is well traced by the internal blue region.}
\label{fig: dm-shield-all}
\end{figure*}

\begin{table}
\centering
\renewcommand{\arraystretch}{1.3}
\tabcolsep=2pt
\footnotesize
\caption{Mean of structural parameters from the last ten snapshots, considering only bound dark matter particles and bound stars.}
\begin{tabular}{l|cccccccc}
\hline\hline             
\multicolumn{1}{c}{ID} &
\multicolumn{1}{c}{$M_{\bullet}$} &
\multicolumn{1}{c}{$M_{\star\bullet}$} &
\multicolumn{1}{c}{$M_{\star}$} &
\multicolumn{1}{c}{$r_{1/2, \bullet}$} &
\multicolumn{1}{c}{$r_{1/2, \star\bullet}$} &
\multicolumn{1}{c}{$r_{1/2, \star}$} &
\multicolumn{1}{c}{$r_{t, \rm GC+DM}$} &
\multicolumn{1}{c}{$r_{t, \rm GC}$}  \\
\multicolumn{1}{c}{} &
\multicolumn{1}{c}{[$10^5 \, \mathrm{M}_{\odot}$]} & 
\multicolumn{1}{c}{[$10^5 \, \mathrm{M}_{\odot}$]} & 
\multicolumn{1}{c}{[$10^5 \, \mathrm{M}_{\odot}$]} & 
\multicolumn{1}{c}{[pc]} & 
\multicolumn{1}{c}{[pc]} & 
\multicolumn{1}{c}{[pc]} &
\multicolumn{1}{c}{[pc]} & 
\multicolumn{1}{c}{[pc]} \\ 
\multicolumn{1}{c}{(1)} &
\multicolumn{1}{c}{(2)} &
\multicolumn{1}{c}{(3)} &
\multicolumn{1}{c}{(4)} &
\multicolumn{1}{c}{(5)} & 
\multicolumn{1}{c}{(6)} & 
\multicolumn{1}{c}{(7)} & 
\multicolumn{1}{c}{(8)} & 
\multicolumn{1}{c}{(9)} \\ 
\hline
GC1 & $  13.7 $ & $  8.7 $ & $  6.1 $ & $  829 $ & $  16.0 $ & $  24.0 $ & $  510 $ & $  600 $ \\
GC2 & $  8.2 $ & $  9.1 $ & $  8.0 $ & $  486 $ & $  12.4 $ & $  19.2 $ & $  179 $ & $  360 $ \\
GC3 & $  15.5 $ & $  9.6 $ & $  8.8 $ & $  191 $ & $  18.3 $ & $  40.9 $ & $  816 $ & $  1223 $ \\
GC4 & $  10.9 $ & $  9.7 $ & $  9.0 $ & $  354 $ & $  12.9 $ & $  23.0 $ & $  483 $ & $  586 $ \\
GC5 & $  9.3 $ & $  9.7 $ & $  8.6 $ & $  181 $ & $  17.7 $ & $  41.8 $ & $  935 $ & $  1157 $ \\
\hline
\end{tabular}
\parbox{\hsize}{Notes: Columns are: (1) Globular cluster ID; (2) Total mass of the surviving dark matter mini-halo (in $10^5$~M$_{\odot}$); (3) Total mass of the surviving globular cluster originally embedded in a dark matter mini-halo (in $10^5$~M$_{\odot}$); (4) Total mass of the surviving globular cluster devoid of dark matter (in $10^5$~M$_{\odot}$); (5) 3D half-number radius of the surviving dark matter mini-halo (in pc); (6) 3D half-number radius of the surviving globular cluster originally embedded in a dark matter mini-halo (in pc); (7) 3D half-number radius of the surviving globular cluster devoid of dark matter (in pc); (8) Tidal radius of the system composed of a globular cluster plus a dark matter mini-halo (in pc); (9) Tidal radius of the globular clustres simulated without a dark matter mini-halo (in pc). Dark matter mini-halos that survived better throughout the simulation have half-mass radii (column 5) smaller than the system's tidal radius (column 8).}
\label{tab: final-params}
\end{table}

\section{Discussion} \label{sec: discussion}

In the previous section, we have consistently showed that, at least during the first orbits of a GC embedded in a DM mini-halo, the DM behaves as a protecting shield that prevents the development of prominent tidal features. Now, we discuss a different aspect, which is under which circumstances this DM shield is effective.

\subsection{Survival of the dark matter mini-halo} \label{ssec: survival-dm}

The DM mini-halo of our simulated GCs was shown to effectively work as a shield against tidal effects. However, not only this mini-halo is gradually removed by being itself stripped beforehand GC stars, as the increase of mass in the GC$+$DM system (with respect to the GCs devoid of DM) also brings the system closer to the centre of the host galaxy\footnote{Through orbital decay from dynamical friction (see Section~\ref{ssec: decay}).}, where tidal effects are stronger. As a result, the DM mini-halo is expected to become negligible with time, so that the question of interest becomes: When is a DM mini-halo effectively disrupted?

The answering of this question starts when comparing the five GCs we simulated, and their different orbital parameters. First, we can separate these five GCs into two categories: (i) The GCs where the DM shield is not effective by the end of our simulation and (ii) GCs whose shield still manages to protect them against strong tidal effects. Looking at Figure~\ref{fig: dm-shield-all}, we can clearly assign GCs~1 and 2 to category (i), while GCs~3, 4 and 5 are better suited to category (ii), with GC~4 approaching to category (i). This assignment is due to the fact that the limit of bound stars in GCs~1 and 2 is comparable or greater than the respective tidal radius of the GC$+$DM system, meaning that the DM shield illustrated in Figure~\ref{fig: tidal-scheme} is no longer well observed. 
We also managed to observe that such disrupted mini-halos presented $r_t > r_{1/2,\bullet}$ at some point of their orbital evolution (Figure~\ref{fig: radii-dm}), where $r_t$ is the tidal radius of the GC$+$DM system, and $r_{1/2,\bullet}$ the half-mass radius of the bound DM particles. This criterion is similar to the relation presented in \cite{Hayashi+03} for the disruption of DM halos.

\begin{figure*}
\centering
\includegraphics[width=0.95\hsize]{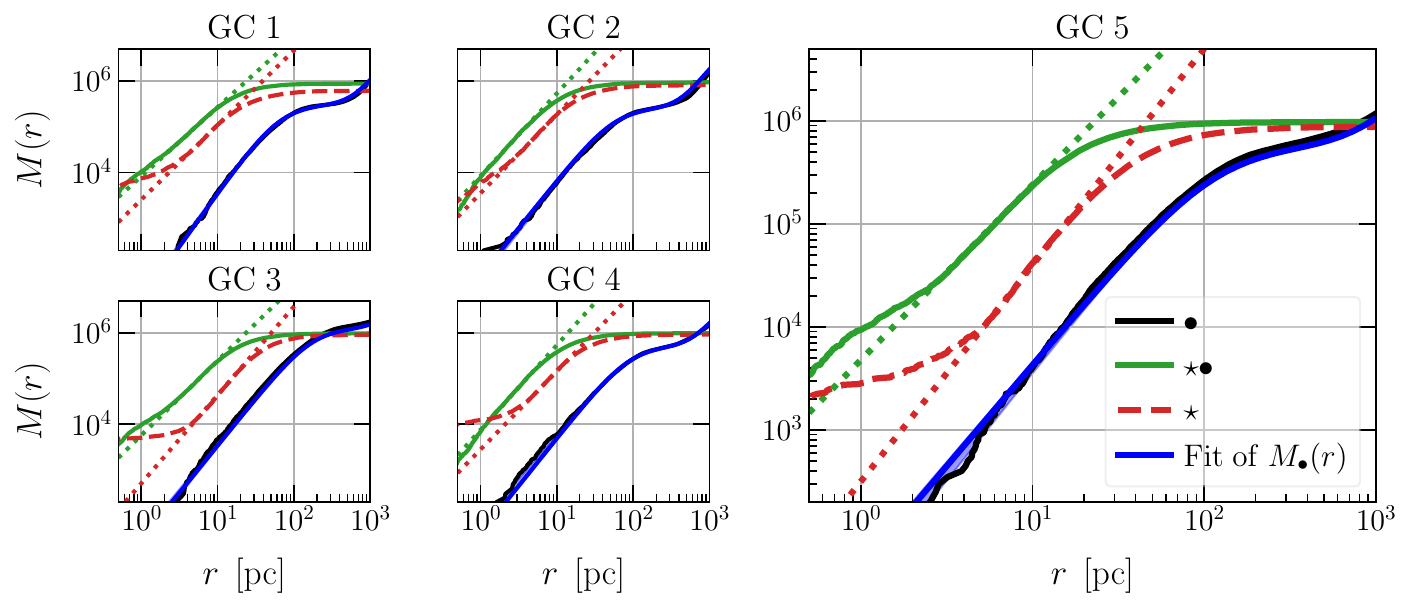}
\caption{Cumulative mass profiles: We display the cumulative mass profiles at the last snapshot of our simulations, for the dark matter particles (\textit{black)}, their embedded stellar component (\textit{green}) and the stellar component from the simulations without dark matter (\textit{dashed red)}. The blue line depicts the MCMC fit of a \protect\cite{Zhao96} $\alpha\beta\gamma$ model to the dark matter mini-halo (see Section~\ref{sssec: ana-dm}), with a fainter blue region encompassing the [2.5,97.5] percentiles. The dotted lines represent local slopes of the stellar mass distribution at the respective regions shown in the plot.}
\label{fig: mass-prof}
\end{figure*}

We argue that the DM loss in GCs~1 and 2 is accelerated by the many passages in pericentre (i.e. relatively small orbital periods), as well as the relatively small orbital radii ($r_{\rm orbit}$) throughout the clusters' orbits. Those factors force these GCs to stay longer next to the centre of the host galaxy, where the tidal interactions are more severe. As a consequence, the DM mini-halos in GCs~1 and 2 suffer from stronger tidal effects and are stripped faster. In Table~\ref{tab: final-params}, we display the mean of the structural parameters of stars and DM particles, over the last ten snapshots of the simulation, and by analysing the difference between columns 5 and 8 (i.e. the half-mass radius of the DM mini-halo, and the tidal radius of the respective system), one can infer that the DM mini-halo survives better to worse, in the following order: GC5 and GC3 similarly, followed by GC4, GC1 and GC2, respectively.

\subsection{Detectability of dark matter}

\subsubsection{On the amount of detectable dark matter}

If some GCs manage to preserve part of their original DM mini-halo, it is important that we understand the observational limitations that may allow us (or not) to detect this DM component. For instance, as the mini-halo mass distribution is more diffuse than the GC stellar component, one might ask if the DM amount in the GC inner regions is significant. For that purpose, we plot in Figure~\ref{fig: mass-prof} the mass profile of different tracers (i.e. DM and GCs, for both scenarios) as a function of distance from the cluster's centre, at the last snapshot of our simulations. 

Figure~\ref{fig: mass-prof} reveals that up to $\sim10$~pc,\footnote{For our GCs embedded in DM, $10$~pc corresponds to roughly 60\% of the half mass radii of the stellar component, for the last snapshot of the simulation.} the mass ratio between the DM component (solid black) and the stellar component of the GCs (solid green) is of the order of $1\%$, meaning that the influence of DM in the internal dynamics of these clusters is mostly negligible\footnote{The plot depicts this behaviour at the end of the simulations, but we stress that throughout the evolution of the cluster, the typical DM cumulative mass remains much lower than the inner GC cumulative mass, given the latter's particularly high values.}.
In the past, many studies \citep{Shin+13,Conroy2011,Ibata+13,Moore1996,Baumgardt2009,Lane2010,Feng2012,Hurst2015} measured very low mass-to-light ratios in GCs, and sometimes used that as an argument to rule out the presence of dark matter in them. We show, in agreement with what has been previously pointed out in \cite{Bromm2002,Mashchenko&Sills&Sills05,Saitoh2006,Bekki2012}, that GCs originally embedded in DM lose much of their initial DM content due to tidal interactions. This naturally leads to mass ratios (stellar to total mass) very close to unity, up to distances as far as available data usually goes.

Figure~\ref{fig: mass-prof} also compares the mass profile of the GCs formed in DM mini-halos and those devoid of it (dashed red). Clearly, clusters with DM have an inner mass much higher than GCs devoid of DM, with GCs~3 and 5, which managed to retain most of their initial DM mini-halo, having inner masses up to ten times higher than their DM-free counterparts. Hence, in addition to denser clusters producing an observational bias where tidal structures are more difficultly detected \citep{Balbinot&Gieles&Gieles18}, we point that the ones that might be formed in DM mini-halos also tend to have at the same time higher inner densities and less prominent tidal structures (see Section~\ref{ssec: tails}), which can render the search for stellar streams in these GCs particularly hard. 

Finally, it is worth mentioning that the outer slopes of the stellar mass profile (dotted lines in Figure~\ref{fig: mass-prof}) do not seem to show a necessary correlation with the analysed scenario (with or without DM), differently than what is shown in \citeauthor{Penarrubia+17} (\citeyear{Penarrubia+17}, figure~2). We associate it once again, as in Section~\ref{ssec: dispersion-stars} for the velocity dispersion profile, with the presence of an intense tidal field (neglected in the former work) that shapes the outer regions of the stellar distribution in a more impactful way than an eventual DM mini-halo.

\subsubsection{Analytical description} \label{sssec: ana-dm}

One of the main ways to look for a DM mini-halo in GCs is to perform mass modelling of observed data, which can be done by solving the Jeans equation \citep{Binney80} for spherical systems with no streaming motions
\begin{equation} 
  \frac{{\rm d}\left (\rho\sigma_r^2\right)}{{\rm d}r} + 2\,\frac{\beta(r)}{
    r}\,\rho(r)\sigma_r^2(r) = -\rho(r) \frac{G\,M_{\rm T}(r)}{r^2} \ ,
\label{eq: jeans}
\end{equation}
where we assume a total mass profile $M_{\rm T}(r)$, a anisotropy profile $\beta(r)$, and a previously determined mass density profile $\rho(r)$ for the kinematic tracers (here, stars). The term $\rho\,\sigma_r^2$ is the dynamical pressure that counteracts gravity. Since the observable tracers are stars, we only need a parametrisation for the DM mass profile in order to model it. This parametric form is then incorporated in the total mass profile $M_{\rm T}(r) = M_{\star\bullet}(r) + M_{\bullet}(r)$. To obtain a suitable mass profile for the DM component, we tested two models, namely the \cite{Kazantzidis+04} model\footnote{This model is
motivated from dynamical simulations of repeated tidal encounters \citep{Kazantzidis+04}.},
\begin{equation} \label{eq: dens-dm-k}
    \rho_{\bullet}(r) = \rho_0 \, r^{-\gamma} \, e^{-r/a}    \  ,
\end{equation}
and the \cite{Zhao96} $\alpha\beta\gamma$ model,
\begin{equation} \label{eq: dens-dm-gp}
        \rho_{\bullet}(r) = \rho_0 \, r^{-\gamma} \, \displaystyle{\left[ 1 + \left( \frac{r}{a} \right)^{\alpha} \right]^{(\gamma - \beta)/\alpha}}    \  ,
\end{equation}
where $\rho_0$ is a scaling factor, $a$ the scale radius, $\gamma \in (0,3)$ an inner slope, $\beta$ an outer slope, which we take to be $5$, such as in the Plummer profile, and $\alpha$ a scaling exponent, taken as $2$, according to the Plummer profile as well.\footnote{When fixing those parameters, the goodness of fit was improved.}
Then, we converted those densities to mass profiles by performing $M_{\bullet}(r) \equiv \int_{0}^{r} 4 \, \pi x^2 \, \rho_{\bullet}(x) \, {\rm d}x$ with \textsc{Mathematica12}. 

Similarly to what is shown in \citeauthor{Mashchenko&Sills&Sills05} (\citeyear{Mashchenko&Sills&Sills05}, figure~7), we noticed that the stripped DM particles from the original mini-halo had a tendency to form a second shell surrounding the remaining mini-halo. Such effect is only noticeable in the DM mass profile beyond a few hundred pc (see Figure~\ref{fig: mass-prof}), but since we choose to analyse the mass profiles up to the outer regions of the clusters, we took that into account by assigning $M_{\bullet}(r) = M_{\bullet,1}(r) + M_{\bullet,2}(r)$, where $M_{\bullet,1}(r)$ is an inner component, most susceptible to be detected or to influence the dynamics of the embedded star cluster, and $M_{\bullet,2}(r)$ is an outer shell, modelled only for the sake of better constraining the outer slope of $M_{\bullet,1}(r)$. This yields, for the \cite{Kazantzidis+04} profile,
\begin{equation} \label{eq: mass-dm-k}
    M_{\bullet,i}(r) = 4 \pi \, a^{3 - \gamma} \, \rho_0 \, \displaystyle{\Gamma{\left(3 - \gamma, \frac{r}{a} \right)}}
\end{equation}
and for the \cite{Zhao96} $\alpha\beta\gamma$ profile,
\begin{equation} \label{eq: mass-dm-gp}
    M_{\bullet,i}(r) = \displaystyle{\frac{4 \pi \, a^{3 - \gamma} \, \rho_0}{3 - \gamma}} \, \displaystyle{\left[ 1 + \left( \frac{a}{r} \right)^{2} \right]^{(\gamma - 3)/2}}
\end{equation}
where we note $\Gamma(b,x) = \int_{0}^{x} t^{b-1} e^{-t} \mathrm{d} t$, which is the lower incomplete gamma function, and $i = \{1,2\}$.

Since we fix the $\alpha\beta$ parameters when using the \cite{Zhao96} $\alpha\beta\gamma$ profile, we end up with the same amount of free parameters than the \cite{Kazantzidis+04} model, which allows us to compare the fits from the two models by simply choosing the one with highest likelihood as our preferred model. Both models performed well, and the preference of likelihoods depended not only on the simulated cluster, but on the considered radial range. In Figure~\ref{fig: mass-prof}, we display in blue the fits of the \cite{Zhao96} $\alpha\beta\gamma$ model, which had an overall better performance for radii smaller than 1~kpc. Indeed, going beyond such distances is not of practical interest when modelling observed data, since at such lengths, the confusion with MW interlopers is considerably damaging for the data set. Hence, for data sets going up to a few tens of pc, Jeans modellers could simply use the \cite{Zhao96} $\alpha\beta\gamma$ model, with fixed $\alpha = 2$ and $\beta = 5$, and a single shell (i.e. $M_{\bullet} = M_{\bullet,1}$) to probe the existence of a DM mini-halo in GCs. We provide in Table~\ref{tab: best-fit-dm}, at the Appendixes, the best fit parameters of our simulated clusters (at $z = 0$, for $M_{\bullet,1}$) in each mass model.
% \pbb{Il faut qu'on discute de ca j'ai pas tout comprise! }

\subsection{Comparison with the literature}

We now compare our study with previous attempts to model GCs in DM mini-halos. Beginning with \cite{Mashchenko&Sills&Sills05}, the authors simulated GCs inside a tidal field, using a $M_{\rm DM} / M_{\rm stars}$ ratio of nearly a hundred. Their simulations used static potentials for the host, which tend to underestimate the impacts of dynamical friction and the trimming of outer structures due to passages in high density regions. Their work showed nevertheless that GCs embedded in DM mini-halos are more resilient to severe tidal stripping, in contrast to DM-free GCs, and also that higher mass to light ratios are found in in the clusters outskirts, while the inner regions keep mass to light ratios similar to the purely baryonic case. They finally proposed that presence of obvious stellar streams is probably the only observational evidence that can reliably rule out a presence of significant amounts of DM in GCs. 

Our conclusions are in general very similar to those from \cite{Mashchenko&Sills&Sills05}, with some important caveats. Indeed, we find that the presence of a DM mini-halo tends to initially protect the GC from tidal effects, but our more suitable handling of dynamical friction from a time-changing potential with live particles suggests that clusters with a DM mini-halo sink faster to the clusters centre, where the tidal field is more intense. As a result, clusters whose orbital parameters tend to naturally place them closer to the host will have this effect strengthened, resulting in a faster disruption of their mini-halo, and further cluster dissolution. Thus, the question if GCs embedded in DM are always more resilient to tidal effects is not as simple, and some of them might actually be disrupted faster, if originally orbiting near the host's centre (see Section~\ref{ssec: survival-dm}). 

As in \cite{Mashchenko&Sills&Sills05}, we also find that the DM mass profile is negligible, with respect to the stellar component, up to at least roughly a half-mass radius from the clusters centre (Figure~\ref{fig: mass-prof}). However, we do not associate obvious stellar streams as the only indicator of a DM-free GC. In fact, it is important to mention that although systems with DM do not form \textit{obvious} stellar streams, they can still form milder and thinner ones before the DM mini-halo is entirely disrupted. Another important DM-free indicators could be a very tangential velocity anisotropy in the outskirts and much higher degrees of rotation, which indirectly leads to higher ellipticities as well. These indicators suggest that the GC stellar component is highly impacted by the host's tidal field, hinting that a DM mini-halo is likely not present at current time\footnote{It is important to notice however, that as pointed in \cite{Mashchenko&Sills&Sills05}, such indicators help to rule out a current DM mini-halo, but not necessarily a DM mini-halo formation scenario, since such mini-halos can be disrupted much before a Hubble time.}. We also propose to use IOM space locations to evaluate if the existence of a DM mini-halo is likely (Figure~\ref{fig: action-angle}), since that clusters formed in this scenario tend move towards regions of lower energy in the $E \times L_{\rm Z}$ diagram, due to dynamical friction, while still keeping in mind that DM-free clusters can eventually form in these regions and remain there during their evolution. Our new analysis of the IOM evolution additionally shows that associations of GCs and accreted satellites through nearby locations in IOM space are not reliable for clusters that were embedded in a DM mini-halo.

Recently, \cite{Penarrubia+17} analysed the effects of isolated GCs embedded in DM mini-halos, with DM to stellar mass ratios of ten and a hundred. Their analyses predicted a flattening or increase of the velocity dispersion profile at outer radii for GCs with DM, along with shallower surface density profiles. They also provided analytical forms for the DM mass profile, which can be used in Bayesian approaches. Although such predictions seem indeed valid for isolated systems, when placing them inside a realistic tidal field, such signatures are mostly erased in the cluster outskirts. We have showed in Section~\ref{ssec: dispersion-stars} that an eventual inflation in the velocity dispersion profile can also happen in DM-free clusters due to tidal heating, sometimes in even more pronounced proportions (see Figure~\ref{fig: evolution-dmax}). When at apocentres though, where tidal interactions are milder, we retrieved similar velocity dispersion shapes than \cite{Penarrubia+17} for GCs embedded in DM mini-halos, since those are better shielded from tides and resemble better the isolated case. We did not find a clear relation between the outer mass slopes of GC with and without DM, likely due to the same arguments above.

As in \cite{Penarrubia+17}, we provide an analytical shape for the mass profiles of DM mini-halos, that can be used in Bayesian analyses, specially in Jeans modelling routines \citep{Read+21}. In our case though, our mass profiles were derived also for partially disrupted mini-halos, such as it is expected from systems evolving in a tidal field. We thus consider our parametric forms better suited to the realistic cases of GCs orbiting the MW or other Local Group galaxies.

Another recent simulation of GCs pertaining to the DM formation scenario was done by \cite{Carlberg&Keating22}. However, the authors do not consider a single mini-halo per cluster such as in our work (or such as in \citealt{Mashchenko&Sills&Sills05,Penarrubia+17}), but they consider the scenario where many GCs are formed inside higher mass sub-halos\footnote{Their minimum sub-halo mass is of $5 \times 10^8 \, \msun$, much higher than the values considered here.}. Although that analysis is also interesting, a direct comparison with our work would be unfair, given such very different initial conditions. 

\subsection{General caveats}

As the simulations we ran required a substantial computer power and running time, we needed to constrain some of our assumptions and exploratory analyses, in order to meet all the numerical convergence tests performed in \cite{Boldrini2020}. In this section, we briefly discuss novel aspects that can be envisaged in future works and how they could affect some of our results, by changing a few of our initial assumptions.

\subsubsection{Mass profiles}

Aiming for a more homogeneous treatment and interpretation, we assume all five GCs and respective mini-halos to have the same mass distributions. Although such assumption is reasonable, given the observed stellar masses of the Fornax dSph GCs \citep{deBoer&Fraser16}, the more general scenario of Galactic GCs for example encompasses a much broader mass and scale radii range \citep{Baumgardt&Hilker18}, which can deliver more different results. Another way to look into this variable is that all of our GCs have different filling factors\footnote{The filling factor is the ratio of of Jacobi to half-mass radii: $r_{\rm J}/r_{\rm h}$}, and one could eventually envisage a numerical simulation where different GCs are formed with a similar filling factor.

In such scenario, clusters and respective mini-halos that lie closer to the host would be more compact than in our simulations, which does seem to happen to some extent in the Milky Way (see GC half-number radii from \citealt{Vitral21} and orbital radii from \citealt{Baumgardt+19}, for example). By being more compact, inner clusters could have their DM shield effective for longer timescales, and thus present milder or no stellar streams at present age, as well as smaller degrees of rotation, ellipticity and higher velocity anisotropies, according to our findings. Similarly, if innermost GCs and respective mini-halos happen to be less massive, dynamical friction would be less prominent, and the GC$+$DM system would sink more slowly to regions of stronger tides. Again, the shielding consequences mentioned above would then last for longer in such clusters.
We thus propose, for instance, that future works could explore a greater variety of GC and mini-halo mass profiles for clusters starting at similar positions with respect to their host.

\subsubsection{Particle masses}

In our simulation, all live objects (i.e. stars and DM particles) are set up with the same mass of $50$~M$_{\odot}$. This numerical recipe helps to avoid possible heating and segregation effects that might affect the long-term internal evolution of the GCs embedded in DM. We thus assume the simplest case of equal particle mass in order not to bias our results in the case of heavier or lighter DM particles, and choose to study rather the external regions of the GCs. Indeed, mass segregation effects are less severe in external cluster regions due to a lower density of particles, which in turn limits the amount of dynamical interactions that lead to energy equipartition.

Nonetheless, if usual DM particles or sub-structures have masses considerably lower or higher than stars, our simulations could then fail to spot possible mass segregation effects specially related to the cluster internal dynamics. Meanwhile, the outer dynamics should not be considerably affected (as explained above), such that our results would remain valid up to a reasonable extent.

\cite{Mashchenko&Sills05_internal} studied the aspects of cluster relaxation, using a particle mass ratio of $m_{\star\bullet} / m_{\bullet} = \{ 0.044, 0.44, 0.88 \}$. They found that because the inner regions of the GC$+$DM system are dominated by stars (as we confirm in this study), the dynamics in the cluster's core is not severely changed, and the impact of DM only starts to be visible at the radius where the amount of DM particles and stars are equivalent. At this point, energy equipartition leads the more massive DM particles to scatter the surrounding stars, creating a cutoff radius similar to what is observed in the density profiles of many Milky Way GCs (e.g. \citealt{Trager+95}). Equivalently, one could expect that less massive DM particles would themselves be scattered and deplete the mini-halo faster, thus decreasing the effectiveness of the shielding mechanism we demonstrated.

\subsubsection{Different orbits}

When comparing simulated GCs with and without a DM mini-halo in a tidal field, the important mass contribution from the mini-halo is such that the orbits of these two scenarios are bound to be different, as a consequence of dynamical friction (e.g. upper panels of Figure~\ref{fig: evolution-anis}). In such setup, it can be sometimes difficult to disentangle which dynamical effects are due to the DM mini-halo, or to the ongoing tidal interactions at a given time step. For that reason, we choose to analyse the evolution of these two scenarios starting from the same distance relative to the host, in an attempt to provide a fairer comparison. This forcibly leads clusters formed inside a DM mini-halo to inner regions of the host's potential, where tidal forces are stronger.

Hence, all indications of milder tidal effects in GCs embedded in DM add robustness to our conclusion of a DM shielding mechanism. This is particularly true for the DM consequences we highlighted on stellar streams, velocity anisotropy, rotation and ellipticity. However, populating our simulations with more GCs would undeniably provide a more complete picture on the distribution of GC positions in IOM space at a given time, for instance. 
Also, by simulating clusters at higher distances from their host, they would suffer less dynamical friction\footnote{This is because at higher distances, there are less host's particles for the clusters to interact with, and thus exchange energy.}, which could allow us to compare velocity dispersion profiles of clusters at more similar orbital radii. Due to the high cost of our simulations, we restrained such study to the five observed clusters in the Fornax dSph, so as to approach a realistic case, but adding similar clusters at a broader range of orbital radii seems to be a tantalising direction for future studies.

\section{Summary \& conclusions} \label{sec: conclusion}

We have used full $N$-body simulations on GPU of the globular cluster system in the Fornax dwarf spheroidal galaxy to probe the differences from globular clusters originally embedded in a dark matter mini-halo, and those devoid of it. For that, we simulated the two cases, where globular clusters have or not an enveloping dark matter mini-halo, using the initial conditions from \cite{Boldrini2020}, where all tracers (stars and dark matter) are assumed as $N$-body particles, for both the globular clusters and the host galaxy. Although the extra mass of the system containing dark matter accelerates orbital decay, sinking the clusters to regions of stronger tidal fields, the same extra mass triggers an increase of the tidal radius, which generally goes well beyond the limiting region of globular cluster bound stars (see Figure~\ref{fig: tidal-scheme}). As a result, this tidal radius growth causes the dark matter mini-halo to work as a protective shield (see Figure~\ref{fig: dm-shield-gc5}), which is itself stripped beforehand globular cluster stars. Below, we summarise some of the main diagnostics from our work.

\begin{itemize}
    \item Globular clusters with a dark matter mini-halo should have, in general, more radial outer velocity anisotropy profiles, throughout all their orbit.
    \item Inflations on the outer velocity dispersion profile are expected in both clusters with and without a dark matter mini-halo, as a result of the host's tidal field.
    \item Globular clusters with a dark matter mini-halo should have smaller degrees of internal rotation, and as a consequence, smaller ellipticities for their stellar distribution.
    \item Globular clusters without a dark matter mini-halo develop more prominent stellar streams, and present a more diffuse stellar distribution.
    \item Due to stronger dynamical friction, clusters originally embedded in a dark matter mini-halo evolve towards considerably lower energy regions of the integrals of motion space (differently from their dark matter-free counterparts), and cannot be reliably associated to previous accretion events due to their fast displacements in such space.
    \item Even clusters that retain large amounts of dark matter tend to have inner densities vastly dominated by baryonic components (up to a factor of a hundred in mass), and dark matter searches should target rather the cluster's outskirts.
    \item We provide parametric mass profiles for dark matter mini-halos that evolved in a tidal field, which is of much usefulness for dark matter mini-halo searches employing Jeans mass-modelling \citep[e.g.][]{Carlberg&Grillmair22}.
\end{itemize}

All in all, our work helps to confirm previous predictions (e.g. the shielding mechanism from dark matter mini-halos), but also challenges a few preceding points. For instance, we can mention the shape of stellar velocity dispersion profiles, which we show to be more dependent on the tidal field rather than the presence of a mini-halo.
In addition, we also point to the long-term survival of clusters embedded in dark matter, which \textit{in some scenarios} can be disrupted faster due to pronounced dynamical friction, since the latter leads them to regions of stronger tidal forces, where the shielding mechanism will vanish in shorter timescales, all while leaving the cluster closer to its host.
Such new predictions are due mostly to our better handling of tidal interactions and dynamical friction, respectively.
We also provide the first direct analysis pertaining to the evolution of the stellar velocity anisotropy, rotation, ellipticity and position in integrals of motion space for the dark matter mini-halo scenario. Finally, we evaluate some of the caveats of our analyses, and point towards novel aspects that can be further explored in more detail with the help of numerical simulations.

\begin{acknowledgements}
We thank the anonymous referee for the constructive report, with many insightful comments that have helped us to improve the quality of our results and clarify some descriptions in the manuscript.
We also thank Geraint Lewis and Amaury Micheli for useful tips concerning the 3D visualisation of stellar streams.
\\
Eduardo Vitral was funded by an AMX doctoral grant from \'Ecole Polytechnique. This work was supported by the EXPLORAGRAM
Inria AeX grant.
\\
We greatly benefited from the public software {\sc Python} \citep{VanRossum09} packages 
{\sc BALRoGO} \citep{Vitral21},
{\sc Scipy} \citep{Jones+01},
{\sc Numpy} \citep{vanderWalt11} and
{\sc Matplotlib} \citep{Hunter07}. We also used 
{\sc GeoGebra} \citep{Hohenwarter02}, as well as the {\sc Spyder} Integrated Development Environment \citep{raybaut2009spyder}.
\end{acknowledgements}

%%%%%%%%%%%%%%%%%%%%%%%%%%%%%%%%%%%%%%%%%%%%%%%%%%
% \section*{Data Availability}

% The data that support the plots within this paper and other findings of this study are available from the corresponding authors upon reasonable request.



\bibliography{src}

%\begin{thebibliography}{99}
%\end{thebibliography}

%@arxiver{evolution_action.pdf,im5_204_zoom.pdf,GC5_shield_evol.pdf}

%%%%%%%%%%%%%%%%%%%%%%%%%%%%%%%%%%%%%%%%%%%%%%%%%%

%%%%%%%%%%%%%%%%% APPENDICES %%%%%%%%%%%%%%%%%%%%%

\begin{appendix}

\section{Velocity anisotropy} \label{app: anis}

The Bayesian computation of the velocity anisotropy in \balrogo\ follows the procedure described in the following. First, we consider the local velocity distribution $h({\bf v},r)$ as a Gaussian product over all spherical coordinates (i.e. $r$, $\theta$, $\phi$), in a similar fashion than done in \mpo\ \citep{Mamon+13,Read+21}
\begin{equation}
  h({\bf v},r) = 
    {\cal G}\left[v_r,\mu_r,\sigma_r(r)\right] \,
    {\cal G}\left[v_\theta,\mu_\theta,\sigma_\theta(r)\right] \,
    {\cal G}\left[v_\phi,\mu_\phi,\sigma_\phi(r)\right] 
\ .
 \label{eq: hvsep}
\end{equation}

Differently than in \mpo, which solves the Jeans equation \citep{Binney&Mamon82} to derive $\sigma_r(r)$, we fit $\sigma_r(r)$ with a generalisation of the parametric Plummer form from \citeauthor{Dejonghe87} (\citeyear{Dejonghe87}, similarly used in \citealt{Vasiliev19c}). 
\begin{equation}
  \sigma_r(r) = \displaystyle{\frac{\sigma_0}{[1 + (r/r_{\sigma})^\alpha]^\gamma}}  \  ,
 \label{sigr}
\end{equation}
where $\sigma_0$ is the radial velocity dispersion at the cluster's centre, $r_{\sigma}$ a characteristic radius, $\alpha$ and $\gamma$ characteristic exponents, which we allow to vary between $\{1, 8\}$, and $\{0.05, 2\}$, respectively. Then, $\sigma_\theta(r)$ and $\sigma_\phi(r)$ can be derived, by assuming spherical symmetry and using the definition of the velocity anisotropy (Eq.~[\ref{eq: anisotropy}]), such that $\sigma^2_\theta(r) = \sigma^2_\phi(r) = \sigma^2_r(r) \, [1 - \beta(r)]$. 
We then use the gOM expression for the anisotropy (Eq.~[\ref{eq: gOM}]) in order to write $\beta(r)$.
Since we do not use the Jeans equation, we can also account for the cluster's systemic motions, and so we assign $\mu_i$ as the mean of the velocities in the $i$ coordinate (differently from \mpo, which fixes $\mu_i = 0$).

\begin{figure}
\centering
\includegraphics[width=0.95\hsize]{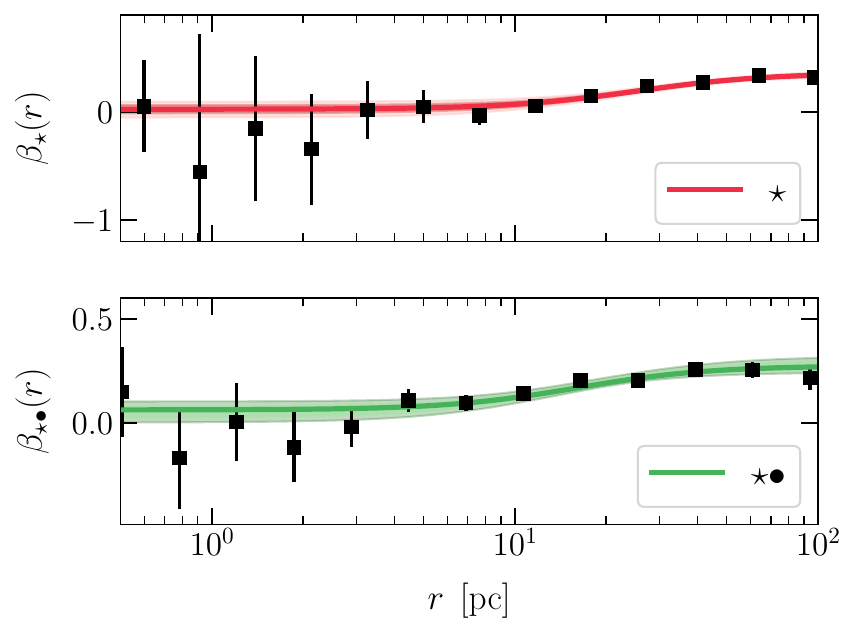}
\caption{Goodness of fit: Bayesian fit of the velocity anisotropy with \balrogo. The \textit{upper plot} shows the case without a dark matter mini-halo, while the \textit{lower plot} shows the case with a dark matter mini-halo. The black dots are the binned values of the empirical velocity anisotropy, logarithmically spaced and with error bars representing the 3-$\sigma$ Poisson error. The solid coloured line (red or green) represent the best likelihood fit of the \protect\cite{Osipkov79,Merritt85} model (Eq.~[\ref{eq: gOM}]), with respective shaded regions encompassing the [$2.5, \, 97.5$] percentiles of the fit.}
\label{fig: anis-goodness}
\end{figure}

Finally, we minimise the logarithm of the likelihood, which is basically the expression from Eq.~\ref{eq: hvsep}, multiplied over each tracer. We do it by using a Markov chain Monte Carlo (MCMC) approach from the \textsc{Python} package \textsc{Emcee} \citep{ForemanMackey+13}, with $10^4$ steps\footnote{The first 5000 were discarded as the initial burn-in phase.} and 15 chains. In Figure~\ref{fig: anis-goodness}, we provide a goodness of fit of the velocity anisotropy of GC5 for the last snapshot from our simulations, for the case with and without a DM mini-halo.

\section{Velocity dispersion}

We describe here how we constructed velocity dispersion profiles for the output of our simulations.

\subsection{Dispersion map} \label{app: disp-map}

The velocity dispersion maps displayed in Figures~\ref{fig: dm-shield-gc5} and \ref{fig: dm-shield-all} was constructed by first binning the projected $X$~vs.~$Y$ map with \textsc{Python}'s \textsc{hexbin} routine, setting the argument \texttt{gridsize}$=100$. For each bin, we then computed the 3D velocity dispersion of the DM component by summing quadratically the velocity dispersion on $X$, $Y$ and $Z$ directions. Whenever there were more than hundred particles inside the bin, we used all the bin's particles, otherwise we completed the sample by picking the closest particles to the bin's centre of mass, until the threshold of a hundred particles was attained.

The process above assured that each bin had a statistically significant number of tracers, which helped us to reach a better spatial resolution, eventually. However, as the results still presented an important amount of statistical noise, we decided to smooth our maps with \textsc{Python}'s \textsc{gaussian\_filter} routine, with the argument \texttt{sigma}$=3$. We finally displayed the outcome of this procedure in a map colour-coded logarithmically from blue (lower dispersion) to red (higher dispersion).

\subsection{Dispersion radial profiles} \label{app: disp-prof}

To build the radial dispersion profiles from Figure~\ref{fig: evolution-dmax}, we used the routines \textsc{angle.cart\_to\_sph} and \textsc{dynamics.dispersion} from the \textsc{BALRoGO} \textsc{Python} package, to first convert the data into spherical coordinates and then compute the velocity dispersion of the radial component, as a function of the distance to the cluster's centre.
The way we set \textsc{BALRoGO} to compute this dispersion is by first dividing the radial extent into thirty equally spaced logarithmic bins and then calculating the respective dispersion and Poisson error associated to it.

Next, it smooths the profile in order to remove statistical noise, by fitting a $10$-th order polynomial to it with the \textsc{numpy.polyfit} routine. 1-$\sigma$ regions were obtained by considering the covariance matrix returned by the same algorithm.
% following the recipe presented in the footnote link\footnote{\url{https://stackoverflow.com/questions/28505008/}}. 
We finally chose to consider the results inside a more restrict spatial extent in order to neglect the potentially bad fits of the polynomial on the borders of our data.

\section{Asymmetric surface density} \label{app: sersic}

Here we describe how the \textsc{BALRoGO}'s \textsc{position.ellipse\_likelihood} routine performs asymmetric S\'ersic fits with $(\alpha,\, \delta)$ data. First, it projects the $(\alpha,\, \delta)$ data according to classical spherical trigonometry relations, translating it so it is centred at the origin:

\begin{subequations}
\begin{flalign}
  \label{x_proj}
    x_{\rm p} &= \cos{\delta} \, \sin{(\alpha - \alpha_{0})} \ ,\\
      \label{y_proj}
    y_{\rm p} &= \sin{\delta} \, \cos{\delta_{0}} - \cos{\delta} \, \sin{\delta_{0}} \, \cos{(\alpha - \alpha_{0})} \ ,
\end{flalign}
\end{subequations}
where $(\alpha_0,\, \delta_0)$ is the centre of the cluster. Next, it rotates the axis so the data can be easily handled:
\begin{subequations}
\begin{flalign}
    x &= x_{\rm p} \, \cos{\theta} + y_{\rm p} \, \sin{\theta} \ , \\
    y &= - x_{\rm p} \, \sin{\theta} + y_{\rm p} \, \cos{\theta} \ ,
\end{flalign}
\end{subequations}
where $\theta$ is the angle between the original reference frame and the new one. With this new set, we are able to define a likelihood function of the stellar distribution as:

\begin{equation}
    \mathcal{L} = \prod_i \, \frac{\Sigma(m)}{N_{\rm tot}} \  .
    \label{eq: lnlike}
\end{equation}
with $m = \sqrt{\left(x/a\right)^2 + \left(y/b\right)^2}$, and where $(a,\, b)$ are the semi-axis of the ellipse. The surface density $\Sigma(m)$ and the number of tracers at infinity $N_{\rm tot}$ are defined such as in \cite{vandeVen&vanderWel&vanderWel21}, so that we have:
\begin{equation}
    \frac{\Sigma(m)}{N_{\rm tot}} = \frac{b_n^{2n} \, \mathrm{exp}\left[ -b_n \, m^{1/n} \right]}{2\pi\, a\,b\,n\,\Gamma{(2n)}}  \  .
    \label{eq: ratio-ell}
\end{equation}

In equation~\ref{eq: ratio-ell}, $n$ is the S\'ersic index, $\Gamma(x)$ is the gamma function of the variable $x$ and the $b_n$ is computed with the precise approximation from \cite{Ciotti&Bertin99}. Thus, the fit finds the parameters which maximise $\mathcal{L}$. 

Finally, to derive statistical errors of our Bayesian estimates, we use \textsc{Python}'s \textsc{numdifftools.Hessian} method to compute the Hessian matrix of the probability distribution function (i.e. eq~[\ref{eq: ratio-ell}]). After, we assign the uncertainties of each parameter as the square root of the respective diagonal position of the inverted Hessian matrix.

\section{Extra material}

\begin{figure}
\centering
\includegraphics[width=0.9\hsize]{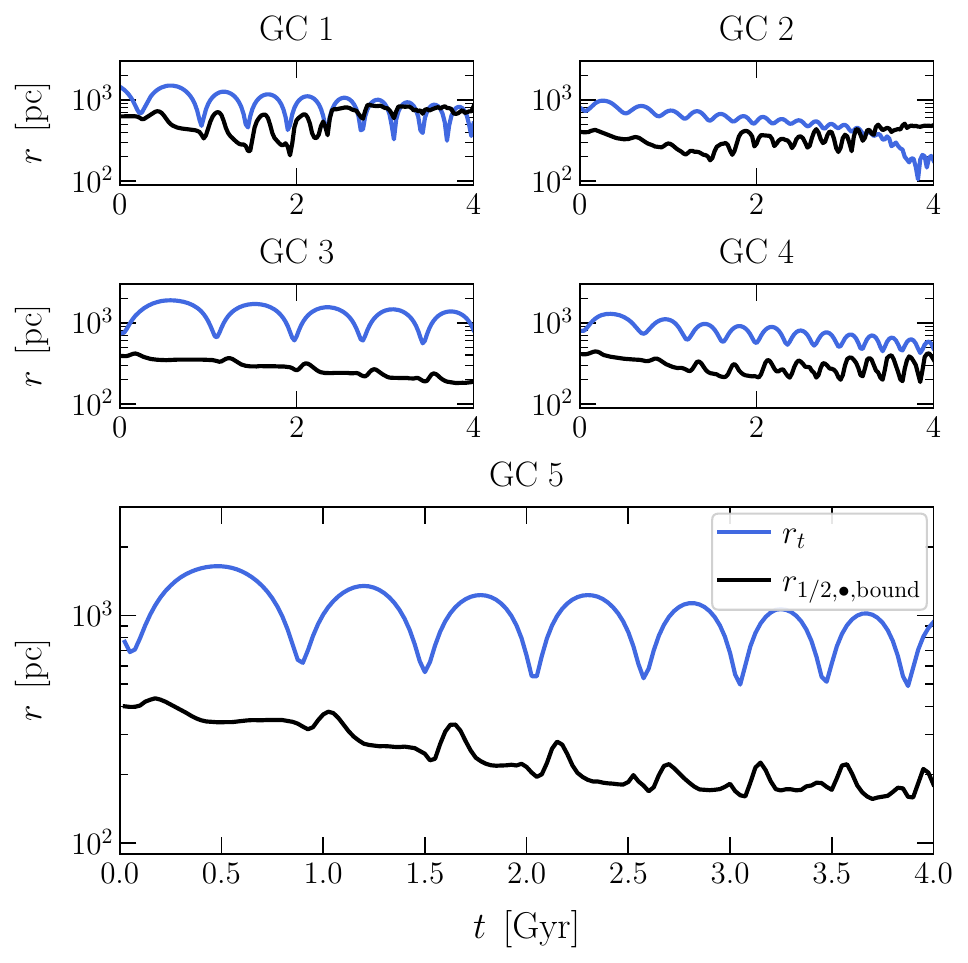}
\caption{Evolution of scale radii: Evolution of the GC$+$DM system tidal radius ($r_t$, \textit{blue}) and of the half-mass radius of the bound dark matter particles ($r_{1/2, \bullet}$, \textit{black}), for each simulated cluster. We observe that clusters who lost most of their dark matter mini-halo (i.e. GC1 and GC2) presented $r_t > r_{1/2, \bullet}$ at some point of their orbital evolution, similarly to the relation for sub-halo disruption from \protect\cite{Hayashi+03}.}
\label{fig: radii-dm}
\end{figure}

\begin{figure}
\centering
\includegraphics[width=0.9\hsize]{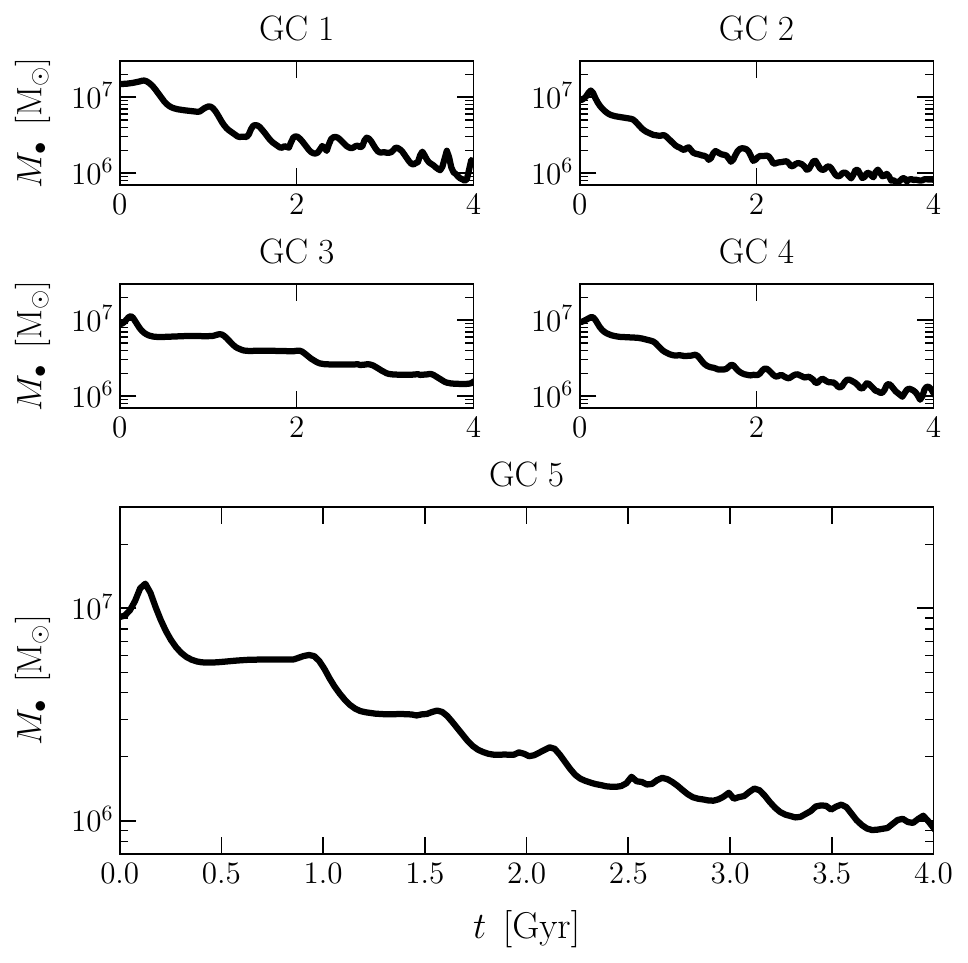}
\caption{Evolution of the dark mass: Evolution of the total mass of bound dark matter particles, for each simulated cluster. Bound particles were defined according to Section~\ref{ssec: data-analysis}.}
\label{fig: mass-dm}
\end{figure}

\begin{figure}
\centering
\includegraphics[width=0.9\hsize]{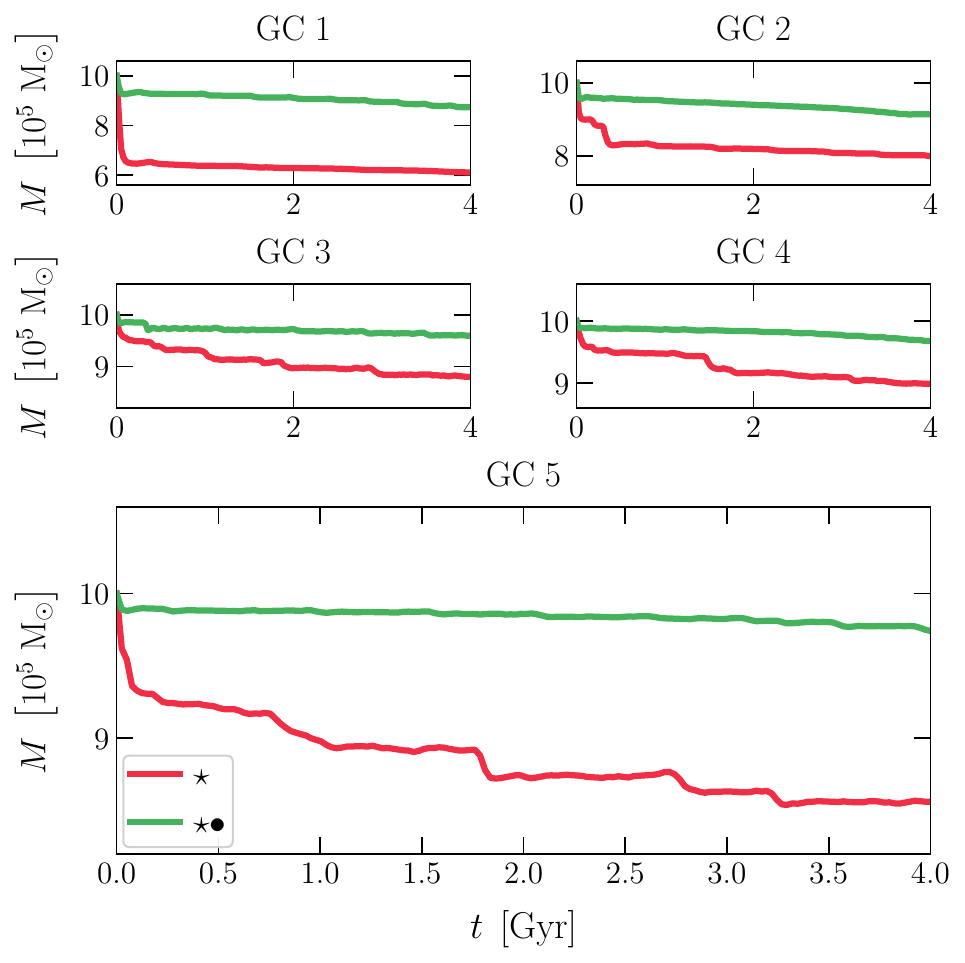}
\caption{Evolution of the stellar mass: Evolution of the total stellar mass of bound globular cluster particles, for each simulated cluster. Clusters originally embedded in a dark matter mini-halo (i.e. $\star\bullet$) are displayed in \textit{green}, while those devoid of the mini-halo (i.e. $\star$) are represented in \textit{red}. Bound particles were defined according to Section~\ref{ssec: data-analysis}.}
\label{fig: mass-gcs}
\end{figure}

\begin{figure}
\centering
\includegraphics[width=0.9\hsize]{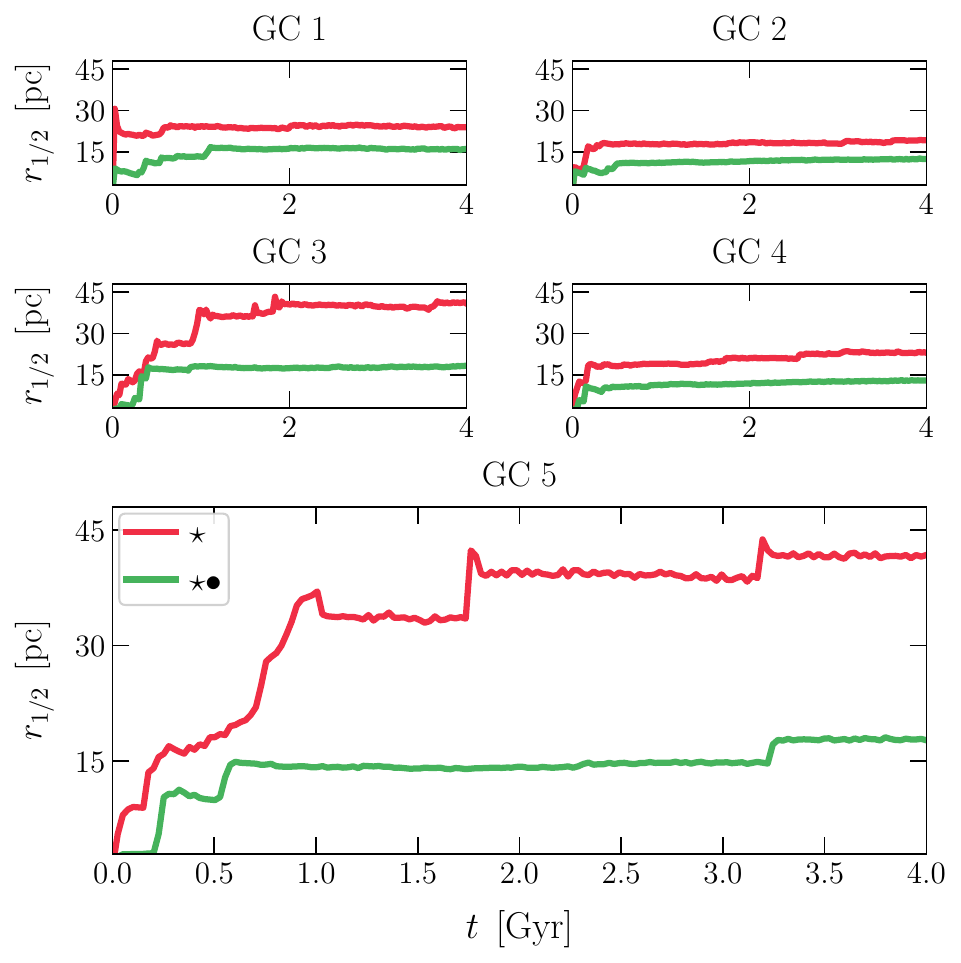}
\caption{Evolution of the half mass radius: Evolution of the half mass radius of bound globular cluster particles, for each simulated cluster. Clusters originally embedded in a dark matter mini-halo (i.e. $\star\bullet$) are displayed in \textit{green}, while those devoid of the mini-halo (i.e. $\star$) are represented in \textit{red}. Bound particles were defined according to Section~\ref{ssec: data-analysis}.}
\label{fig: hmr-gcs}
\end{figure}

\begin{table}
\centering
\renewcommand{\arraystretch}{1.0}
\tabcolsep=6pt
\footnotesize
\caption{Best fit parameters.}
\begin{tabular}{l|rrrrrr}
\hline
\multicolumn{1}{c}{} &
\multicolumn{1}{c}{$a_{\rm Z}$} & 
\multicolumn{1}{c}{$\gamma_{\rm Z}$} & 
\multicolumn{1}{c}{$f_{\rm Z}$} &
\multicolumn{1}{c}{$a_{\rm K}$} & 
\multicolumn{1}{c}{$\gamma_{\rm K}$} & 
\multicolumn{1}{c}{$f_{\rm K}$}   \\
\multicolumn{1}{c}{} &
\multicolumn{1}{c}{[pc]} &
\multicolumn{1}{c}{} &
\multicolumn{1}{c}{} & 
\multicolumn{1}{c}{[pc]} & 
\multicolumn{1}{c}{} & 
\multicolumn{1}{c}{} \\
\hline
GC~1  &   $ 67.2$  &  $ 0.66$  &  $ 0.03$  &  $  28.9$  &  $  0.27$  &  $  0.03$  \\
GC~2  &   $ 58.1$  &  $ 0.93$  &  $ 0.03$  &  $  27.3$  &  $  0.57$  &  $  0.02$  \\
GC~3  &   $ 175.6$  &  $ 0.94$  &  $ 0.13$  &  $  101.0$  &  $  0.94$  &  $  0.13$  \\
GC~4  &   $ 92.0$  &  $ 0.92$  &  $ 0.05$  &  $  46.1$  &  $  0.82$  &  $  0.04$  \\
GC~5  &   $ 119.8$  &  $ 1.03$  &  $ 0.06$  &  $  66.7$  &  $  0.96$  &  $  0.06$  \\
\hline
\end{tabular}
\parbox{\hsize}{Notes: Best likelihood parameters from our mass profile fits for each simulated globular cluster at $z = 0$. The first three columns yield the fits when using a \protect\cite{Zhao96} $\alpha\beta\gamma$ model with fixed $\alpha = 2$ and $\beta = 5$ (as in the classic Plummer model) for better convergence, while the last three columns yield the fits when adopting a \protect\cite{Kazantzidis+04} model. Each three columns display the scale radius $a$, inner slope $\gamma$ and mass fraction $f$ relative to the first dark matter shell explained in Section~\ref{sssec: ana-dm}, defined as $M_{\bullet, 1}$. For details, see Eqs.~\ref{eq: dens-dm-k} and \ref{eq: dens-dm-gp}.}
\label{tab: best-fit-dm}
\end{table}

\end{appendix}

\end{document}